\documentclass[final]{siamltex}

\usepackage{amsmath}
\usepackage{amssymb}
\usepackage{graphicx}
\usepackage{psfrag} 
\usepackage{dcolumn}
\usepackage{bm}
\usepackage[latin1]{inputenc}
\usepackage{texdraw}

\newcommand{\vn}{\mathbf{n}}        
\newcommand{\eps}{\varepsilon}  
\newcommand{\Lap}{\Delta}               
\newcommand{\spr}{s_{\rm pr}}               
\newcommand{\x}{\bar{x}}                
\newcommand{\z}{\bar{z}}                
\newcommand{\p}[1]{\partial_{#1}}   
\newcommand{\BO}[1]{\mathcal{O}\left(#1\right)}     
\newcommand{\so}[1]{o\left(#1\right)}       
\newcommand{\lp}{\left(}                
\newcommand{\rp}{\right)}               
\newcommand{\qlp}{\left[}               
\newcommand{\qrp}{\right]}          
\newcommand{\fetawp}[1]{\frac{#1}{\eta}}                            

\newcommand{\expeps}{\er^{-\sqrt{3}\,\omega z/\xi}}

\font\svfilt=msbm7
\def\svfilS{\hbox{\svfilt S}}
\def\bv#1{\hbox{$\mathbf{#1}$}}
\def\tr{\mathop{\rm tr}\nolimits}
\def\er{{\rm e}}
\def\mr{{\rm m}}

\title{Boundary-roughness effects in nematic
liquid crystals}
\author{Paolo Biscari and Stefano Turzi\thanks{Dipartimento di Matematica,
Politecnico di Milano, Piazza Leonardo da Vinci 32, 20133 Milano
(Italy). E-mail: \texttt{paolo.biscari@polimi.it,
stefano.turzi@mate.polimi.it}} }

\begin{document}

\maketitle

\begin{abstract}
We study the equilibrium configuration of a nematic liquid crystal
bounded by a rough surface. The wrinkling of the surface induces a
partial melting in the degree of orientation. This softened region
penetrates the bulk up to a length scale which turns out to coincide
with the characteristic wave length of the corrugation. Within the
boundary layer where the nematic degree of orientation decreases,
the tilt angle steepens and gives rise to a nontrivial structure,
that may be interpreted in terms of an effective weak anchoring
potential. We determine how the effective surface extrapolation
length is related to the microscopic anchoring parameters. We also
analyze the crucial role played by the boundary conditions assumed
on the degree of orientation. Quite different features emerge
depending on whether they are Neumann- or Dirichlet-like. These
features may be useful to ascertain experimentally how the degree of
orientation interacts with an external boundary.
\end{abstract}



\pagestyle{myheadings} \thispagestyle{plain} \markboth{P.\ BISCARI
AND S.\ TURZI}{BOUNDARY-ROUGHNESS EFFECTS IN NEMATICS}


Nematic liquid crystals are fluid aggregates of elongated molecules.
When the nematic rods interact with an external surface, the elastic
strain energy induces them to align parallel to the unit normal,
even if the surface is not perfectly flat \cite{berr1972}. Recent
experimental observations confirm that the surface alignment of the
nematic director is completely determined by the roughness-induced
surface anisotropy \cite{kuki2005}. Further analytical calculations,
performed within the classical Frank model with unequal elastic
constants, have detected the bulk effects induced by a
periodically-modeled external boundary \cite{babe2005}.

A crucial effect, still related to surface roughness, escapes the
framework of Frank theory, where the only order parameter is the
director. Indeed, it is physically intuitive that nematic molecules
will disorder if we force them to follow a rapidly varying boundary
condition. This \emph{surface melting\/} was first experimentally
detected in \cite{faga1985,barb1990}. Recent experimental
observations have also measured a boundary-layer structure in the
degree of orientation \cite{xuto2004}. The surface melting has been
confirmed by approximated analytical solutions \cite{barb1991},
numerical calculations \cite{skac1998,moce1999}, and molecular
Monte-Carlo simulations \cite{cheu2005}.

The combined effect of a rapidly-varying director anchoring and
surface melting gives rise to an effective weak-anchoring effect
that was first observed in \cite{sasa1992}. The problem of relating
the effective anchoring extrapolation length to the microscopic
roughness parameters has been studied in several theoretical papers,
all framed within the Frank theory
\cite{evan1993,evan1994,alex1994}. This observation is of basic
significance, since weak anchoring potentials have proven to
influence deeply all nematic phenomena, beginning with Freedericksz
transitions \cite{stri1988,Virga,napo2006}. Indeed, several
theoretical studies have already determined the influence on
anchoring energies of the presence of permanent surface dipoles
\cite{ossl1997} or diluted surface potentials
\cite{sovi2000,sovidu2000}.

In this paper we analyze in analytical detail the boundary-layer
structure induced by a rough surface which bounds a nematic liquid
crystal. We frame within the Landau-de Gennes order-tensor theory,
to be able to detect the effects on both the director and the degree
of orientation. Our results confirm the surface melting already
obtained in \cite{barb1991}, but allow us to detect new phenomena.
First, the nematic director steepens close to the boundary, so
giving rise to an effective weak anchoring potential, that turns out
to be deeply related to the surface-melting effect, and thus can be
correctly handled only within the order-tensor theory. Furthermore,
the boundary layers display a strong dependence on the type of
boundary conditions imposed on the degree of orientation. Indeed,
the orders of magnitude of all the expected effects depend on
whether the boundary conditions are Neumann- or Dirichlet-like. We
discuss how these effects may help in ascertaining in experiments
how the mesoscopic parameter, which measures he degree of order,
interacts with an external surface.

The paper is organized as follows. In Section \ref{sec:ec} we
present the model, we set the geometry of a roughly-bounded sample,
and derive the Euler-Lagrange partial differential equations that
determine the equilibrium configurations. In Section \ref{sec:res}
we perform the perturbative two-scales analysis that provides all
the analytical details of the boundary-layer structure. In Section
\ref{sec:effwa} we solve an effective problem, in which the rough
surface is replaced by a weak-anchoring potential. The concluding
Section \ref{sec:disc} compares our outcomes with the effective
results of Section \ref{sec:effwa}, and draws the conclusions.

\section{Equilibrium configurations}\label{sec:ec}

The degree of order decrease has been recognized by many authors as
a crucial effect of surface roughness \cite{barb1991,moce1999}. We
thus describe nematic configurations in the framework of the
Landau-de Gennes \bv{Q}-tensor theory \cite{dgpr}. The order tensor
is defined as the deviatoric part of the second-moment of the
probability distribution of molecular orientations:
\begin{equation}
\bv{Q}(\bv{r}):=\int_{\svfilS^2} (\bv{m}\otimes\bv{m})\,f_r(\bv{m})
\, da-\frac{1}{3}\,\bv{I}\;, \label{defq}
\end{equation}
where \bv{I} denotes the identity tensor. \bv{Q} is a second-order
traceless symmetric tensor, with ${\rm sp}\, \bv{Q}\subset
\big[-\frac{1}{3},\frac{2}{3}\big]$ \cite{Virga}.

In order to keep computations simple, we adopt the one-constant
approximation for the elastic part of the free energy functional
\begin{equation}
f_{\rm el}[\bv{Q}]=\frac{1}{2}\,K \,|\nabla\bv{Q}|^2\,,
\label{sigel}
\end{equation}
where $K$ is an average elastic constant. We stress, however, that
it is straightforward to generalize all what follows to take into
account unequal material elastic constants.

The free-energy functional includes the Landau-de Gennes
thermodynamic potential as well
\begin{equation}
f_{\rm LdG}(\bv{Q})=A\tr\bv{Q}^2-B\tr\bv{Q}^3+C\tr\bv{Q}^4\;.
\label{sigldg}
\end{equation}
The material parameter $A$ depends on the temperature, and in
particular it becomes negative deep in the nematic phase. On the
contrary, $B,C$ can be assumed to be positive and
temperature-independent. The potential (\ref{sigldg}) strongly
favors uniaxial phases, in which at least two of the three
eigenvalues of \bv{Q} coincide. In fact, \bv{Q} is expected to
abandon uniaxiality mainly close to director singularities
\cite{shsl1987,bigu1997,bisl2003}. We will not deal with any defect
structure. Thus, though the uniaxiality constraint is not essential
for our purposes, we follow the attitude of avoiding unnecessary
complications \cite{bica1994,bipe1997}, and restrict our attention
to uniaxial states
\begin{equation}
\bv{Q}(\bv{r})=s(\bv{r})\left(\bv{n}(\bv{r})\otimes\bv{n}(\bv{r})-
\frac{1}{3}\,\bv{I}\right)\;.
\label{quni}
\end{equation}
The scalar $s\in\left[-\frac{1}{2},1\right]$  and the unit vector
\bv{n} are respectively the \emph{degree of orientation\/} and the
\emph{director\/}. With the aid of (\ref{quni}), the potentials
(\ref{sigel}),(\ref{sigldg}) can be written as
\begin{equation}
f_{\rm el}[s,\bv{n}]=K \left( s^2 |\nabla \bv{n}|^2
+\textstyle\frac{1}{3}|\nabla s|^2\right)\quad{\rm and}\quad f_{\rm
LdG}(s)=\textstyle\frac{2}{3}A\,s^2-\frac{2}{9}B\,s^3
+\frac{2}{9}C\,s^4\;. \label{potsn}
\end{equation}
When $A\leq B^2/(12C)$, the absolute minimum of the function $f_{\rm
LdG}(s)$ occurs at the \emph{preferred degree of orientation}
\begin{equation}
\spr:=\frac{3B+\sqrt{9B^2-96AC}}{8C}>0\;. \label{spr}
\end{equation}
In order to gain physical interpretation of the results, we also
introduce the \emph{nematic coherence length} $\xi$ and the
dimensionless (positive) parameter $\omega$, defined as
\begin{equation}
\xi^2:=\frac{9K}{C}\qquad {\rm and}\qquad
\omega^2:=\frac{2}{3}\,(\spr B-4A)\;. \label{xinw}
\end{equation}
The nematic coherence length compares the strength of the elastic
and thermodynamic contributions to the free energy functional. We
will show below that it characterizes the size of the domains where
the degree of orientation may abandon its preferred value $\spr$.
The number $\omega$ depends on the depth of the potential well
associated with $\spr$. Indeed, it is defined in such a way that
$f''_{\rm LdG}(\spr)=\omega^2/\xi^2$.

By using (\ref{spr}),(\ref{xinw}) we write the bulk free-energy
density $f_{\rm b}:=f_{\rm el}+f_{\rm LdG}$ as
\begin{equation}
\frac{f_{\rm b}[s,\bv{n}]}{K}=  s^2 |\nabla \bv{n}|^2 +
\frac{1}{3}|\nabla s|^2+\frac{1}{\xi^2} \left( s^4 - \frac{4}{3}s^3
\left(2 \spr - \frac{\omega^2}{\spr}\right) + 2s^2(\spr^2 -
\omega^2)\right). \label{fb}
\end{equation}

\subsection{Modelling a rough surface}

We aim at analyzing the effects that a rough boundary induces in a
nematic liquid crystal. Once again, we try to keep our analysis as
simple as possible, while still catching the essential features. We
thus follow \emph{e.g.\/} \cite{evan1993} in modeling roughness by
imposing a sinusoidally-perturbed homeotropic anchoring condition on
a flat surface. The amplitude and the wave length characterizing the
perturbation will be the crucial parameters in our results.

We focus attention on a thin boundary layer, attached to the
external surface. Consequently, we disregard the detailed structure
of the bulk equilibrium configuration, that will only enter our
results as asymptotic \emph{outer\/} solution for the surface
boundary layer. We introduce a Cartesian frame of reference
$\{\bv{e}_x,\bv{e}_y,\bv{e}_z\}$, and assume that the nematic
spreads in the whole half-space $\mathcal{B}=\{z\geq 0\}$. We
further simplify the geometry by assuming that
$\bv{n}(\bv{r})=\sin\theta(\bv{r})\,\bv{e}_x
+\cos\theta(\bv{r})\,\bv{e}_z$ and that the asymptotic bulk
configuration depends only on $z$
\begin{equation}
\theta(\bv{r})\approx\theta_{\rm b}(z) \qquad {\rm as} \quad
z\to+\infty\;. \label{thli}
\end{equation}
In the presence of strong homeotropic anchoring on a flat surface,
the boundary condition to be imposed on the director would be
$\theta^{\rm (flat)}(x,y,0)=0\,$. On the contrary, we will require
\begin{equation}
\theta(x,y,0)=\Delta\cos\frac{x}{\eta}\;. \label{bc}
\end{equation}
The boundary condition (\ref{bc}) mimics the rugosity of the
external surface by introducing two new parameters: the
(dimensionless) roughness amplitude $\Delta$ and the roughness
length $\eta$. We remark that the oscillation rate increases as
$\eta\to0^+$, while all roughness effects are expected to vanish in
the limit $\Delta\to 0^+$. The requirements (\ref{thli}),(\ref{bc})
imply that the free-energy minimizer will not exhibit any dependence
on the transverse $y$-coordinate, so that we will henceforth
restrict attention to the dependence on the coordinates $(x,z)$.

It is more complex to ascertain the correct type of boundary
conditions which are to be imposed on the degree of orientation $s$.
From the mathematical point of view, it would be natural to imitate
the (Dirichlet) strong anchoring imposed on the director, and thus
set $s(x,y,0)$ to be equal to some fixed boundary value $\tilde s$.
Nevertheless, while it is well-established that we can induce an
easy axis for the director on an external boundary, it is
questionable whether we may fix the value of a mesoscopic parameter,
that measures the degree of order in a distribution. From the
physical point of view it would appear then more natural to impose
(Neumann) free boundary conditions on the degree of orientation,
leaving to the thermodynamic potential (\ref{sigldg}) the assignment
of inducing the preferred value $\spr$ in the bulk ($z\to\infty$).
To be safe, both possibilities (Dirichlet and Neumann) will be
analyzed in Section \ref{sec:res}.

\subsection{Euler-Lagrange equations}

Once we consider that $|\nabla \vn |^2 = |\nabla \theta |^2$, it is
easy to derive the Euler-Lagrange partial differential equations
associated with the functional (\ref{fb}). They read:
\begin{equation}
\label{EL0102}
 s^2 \Lap \theta + 2\, s\, \nabla s \cdot \nabla \theta = 0
 \qquad{\rm and}\qquad
\Lap s - 3\, s\,|\nabla \theta|^2 - 3\, \frac{\sigma(s)}{\xi^2}=0\,,
\end{equation}
where
\begin{equation}
\sigma(s):=s (s-\spr) \left(s- \spr +
\frac{\omega^2}{\spr}\right)\;.\label{therpot}
\end{equation}
Since the boundary conditions (\ref{bc}) are $x$-periodic, with a
period of $2\pi\eta$, we look for solutions of (\ref{EL0102}) in
$C^2_{2\pi\eta}$ (the space of $C^2$-functions, $2\pi\eta$-periodic
in the $x$-direction). To complete the differential system
(\ref{bc}), in \S\ref{subsec:neu} we will require
\begin{equation}
\begin{cases}
\theta(x,0)=\Delta \cos \displaystyle\frac{x}{\eta}\\
\displaystyle\frac{\partial s}{\partial z}(x,0)=0
\end{cases}
\qquad{\rm and}\qquad
\begin{cases}
\theta(x,z)\approx\theta_{\rm b}(z) \\
s(x,z)\approx\spr
\end{cases}
\quad{\rm as}\ z\to\infty \label{bcneu}
\end{equation}
while in \S\ref{subsec:dir} we will choose
\begin{equation}
\begin{cases}
\theta(x,0)=\Delta \cos \displaystyle\frac{x}{\eta}\\
s(x,0)=\tilde s
\end{cases}
\qquad{\rm and}\qquad
\begin{cases}
\theta(x,z)\approx\theta_{\rm b}(z) \\
s(x,z)\approx\spr
\end{cases}
\quad{\rm as}\ z\to\infty \label{bcdir}
\end{equation}

\section{Two-scales analysis}\label{sec:res}

Before proceeding with the perturbation analysis of the differential
equations, we state them in dimensionless form. It will turn out
that the correct scaling is obtained by measuring lengths in
$\eta$-units, so that we introduce the new dimensionless coordinates
$\x=x/\eta$, $\z=z/\eta$, and define the dimensionless parameter
$\eps\,=\,\xi/\eta$. Equations (\ref{EL0102}) become thus
\begin{equation}
\label{EL12} s^2 \Lap \theta + 2 s\, \nabla s \cdot \nabla \theta =
0 \qquad{\rm and}\qquad  \eps^2\Lap s - 3\eps^2 s\,|\nabla \theta|^2
- 3 \sigma(s)=0\,,
\end{equation}
where both the gradient and the laplacian are now to be intended
with respect to the scaled variables. The nematic coherence length
is usually much smaller than all other characteristic lengths.
Consequently, we will look for uniformly asymptotic solutions to
(\ref{EL12}), by treating $\eps$ as a small parameter. In this
limit, equation (\ref{EL12}$)_2$ is singular, so that a regular
asymptotic expansion would not provide a uniform approximation of
the solution. Indeed, the small parameter $\eps$ multiplies the
highest derivative, so that we may expect the solution to have a
steep behavior in a layer of thickness $\delta$ (to be determined),
close to the boundary $z=0$. We refer the reader to the books
\cite{Holmes,Bender,Murdock,Smith} for the details of the singular
perturbation theory we will apply henceforth and, in particular, for
the technique of the two-scales method which directly yields a
uniform approximation of the solution.

A standard dominant balance argument (that requires to introduce a
stretched variable $Z=\z/\delta$) allows to recognize that the
boundary layer thickness is $\delta=\eps$. We then introduce the
\emph{fast\/} variable $Z=\z/\eps$. The two-scales chain rule
requires to replace $\p{\z}$ by $\big(\p{\z}+\eps^{-1}\p{Z}\big)$,
and equations (\ref{EL12}) take the form (when $s\neq 0$)
\begin{align}
s \left(\eps^2\theta_{,\x \x} +\eps^2\theta_{,\z \z}
+2\eps\theta_{,\z Z} +\theta_{,Z Z} \right) +
2\eps^2s_{,\x}\theta_{,\x}+ 2(\eps s_{,\z}+s_{,Z})(\eps\theta_{,\z}+
\theta_{,Z})&=0 \label{ELtt}\\
\eps^2s_{,\x \x}+\eps^2s_{,\z \z}+2\eps s_{,\z Z}+s_{,Z Z} - 3
s\left[\eps^2(\theta_{,\x})^2+(\eps\theta_{,\z}+\theta_{,Z})^2
\right] - 3\sigma(s)&=0\label{ELs}
\end{align}
where a comma denotes differentiation with respect to the indicated
variable. In agreement with the two-scales method, $\theta$ and $s$
are now to be intended as $\theta(\x,\z,Z)$ and $s(\x,\z,Z)$, that
is functions of $\x,\z$ and $Z$ regarded as \emph{independent}
variables. It will be only at the very end of our calculations that
we will recast the connection between $\z$ and $Z$: $Z=\z/\eps$. We
seek for solutions which may be given the formal expansions
\begin{align}
\label{exp_theta}
\theta(\x,\z,Z)&=\theta_0(\x,\z,Z)+\eps \theta_1(\x,\z,Z)+ \eps^2
\theta_2(\x,\z,Z)+ O(\eps^3)\\
\label{exp_s} s(\x,\z,Z)&=s_0(\x,\z,Z)+\eps s_1(\x,\z,Z)+ \eps^2
s_2(\x,\z,Z)+O(\eps^3)\;.
\end{align}
If we insert (\ref{exp_theta})-(\ref{exp_s}) in
(\ref{ELtt})-(\ref{ELs}), we obtain the following sequence of
differential equations to $\BO{1}$, $\BO{\eps}$, and $\BO{\eps^2}$
\begin{align}
&\begin{cases} \frac{1}{s_0} \left(s_{0}^2
\theta_{0,Z}\right)_{,Z}=0 \\
 s_{0,ZZ} - 3 s_0(\theta_{0,Z})^2 - 3 \sigma(s_0)=0
\end{cases}\label{O1} \\
&\begin{cases} \frac{1}{s_0} \left(s_{0}^2 \theta_{1,Z}\right)_{,Z}
+ \frac{1}{s_1} \left(s_{1}^2 \theta_{0,Z} \right)_{,Z} =
-2\left(s_0
\theta_{0,Z}\right)_{,\z} -2s_{0,Z}\theta_{0,\z} \\
s_{1,Z Z} - 6 s_0 \theta_{0,Z}\theta_{1,Z} - 3s_1 \left(
\sigma^{\prime}(s_0)+(\theta_{0,Z})^2 \right) =6 s_0
\theta_{0,Z}\theta_{0,\z} - 2 s_{0,\z Z}\end{cases}\label{Oe}\\
& \begin{cases} \frac{1}{s_0} (s_{0}^2 \theta_{2,Z} )_{,Z} +
\frac{1}{s_2}(s_{2}^2\theta_{0,Z})_{,Z}= -\frac{1}{s_1}
(s_{1}^2\theta_{1,Z} )_{,Z} - \frac{1}{s_0} (s_{0}^2\theta_{0,\z}
)_{,\z} -
\frac{1}{s_0} (s_{0}^2\theta_{0,\x} )_{,\x}\\
\phantom{\frac{1}{s_0} \left(s_{0}^2 \theta_{2,Z} \right)_{,Z} }-
2\left(s_0\theta_{1,Z}\right)_{,\z} -
2\left(s_1\theta_{0,Z}\right)_{,\z} -
2s_{1,Z}\theta_{0,\z} - 2s_{0,Z}\theta_{1,\z}\\
s_{2,Z Z} - 3s_2\left[\sigma^{\prime}(s_0) + (\theta_{0,Z})^2
\right]-6s_0\theta_{0,Z}\theta_{2,Z}=
\frac{3}{2}s_{1}^2\sigma^{\prime\prime}(s_0) \\
\phantom{s_{2,Z Z} } + 3s_0\left[\left(\theta_{0,\z} +
\theta_{1,Z}\right)^2 +
(\theta_{0,x})^2 \right]\\
\phantom{s_{2,Z Z} }+
6\theta_{0,Z}\left(s_1\theta_{1,Z}+s_1\theta_{0,\z}+s_0\theta_{1,\z}\right)
- 2s_{1,\z Z} - s_{0,\z \z} - s_{0,x x}\;. \end{cases} \label{Oe2}
\end{align}
Analogous equations can be easily derived at any desired order. For
any $n\geq 1$, the differential system obtained at $\BO{\eps^n}$ is
linear in the unknowns $\theta_n$,$s_n$, and may be solved
analytically. By virtue of the multiscale method, we find the
correct dependence on $\z,Z$ by requiring that all $s_n,\theta_n$
are uniformly bounded as $\eps \rightarrow 0^+$ for expanding
intervals of the type $0 \leq Z \leq Z^{*}/\eps$, for a suitable
positive constant $Z^{*}$. In most practical cases this requirement
is equivalent to asking the removal of secular terms (\emph{i.e.\/}
terms that diverge as $Z \rightarrow +\infty$).

\subsection{Free surface degree of orientation}\label{subsec:neu}

In terms of the scaled variables, the boundary conditions
(\ref{bcneu}) require
\begin{equation}
\begin{cases}
\theta(\x,0)=\Delta \cos \x\\
s_{,\z}(x,0)=0
\end{cases}
\qquad{\rm and}\qquad
\begin{cases}
\theta(\x,\z)\approx\theta_{\rm b}(\eta\z) \\
s(\x,\z)\approx\spr
\end{cases}
\quad{\rm when}\ \z\gg\eta\;.\label{bcneu1}
\end{equation}
We introduce $\mr:=\theta'_{\rm b}(0)$, the derivative of the
asymptotic solution at $z=0$, since it will play an important role
in the following discussion. The leading solutions in expansions
(\ref{exp_theta}),(\ref{exp_s}) are
\begin{equation}
s_0(x,z)=\spr \qquad{\rm and}\qquad \theta_0(x,z)=\mr \,z + \Delta
\er^{-z/\eta}\cos\fetawp{x}\;.\label{sol1}
\end{equation}
Higher order asymptotic solutions are gathered by means of laborious
but straightforward calculations. After recasting the solutions in
terms of the dimensional variables $x = \eta\,\bar{x}$ and $z =
\eta\,\bar{z}$, we find
\begin{align}
 s(x,z) & =\spr - \frac{\spr \xi^2}{\omega^2}\left(
\mr^2 - \frac{2\,\mr\,\Delta}{\eta}\,\er^{-z/\eta}\,\cos\fetawp{x} +
\frac{\Delta^2}{\eta^2}\,\er^{-2z/\eta}\right)\nonumber\\
\label{s_sol_neu}&+ \frac{2\spr\,\xi^3}{\sqrt{3}\,\omega^3}\,
\er^{-\sqrt{3}\omega z/\xi} \left(\frac{\Delta^2}{\eta^3} -
\frac{\mr\,\Delta}{\eta^2}\cos\fetawp{x}\right) +\BO{\eps^4}\quad
{\rm and}\\
\label{theta_sol_neu} \theta(x,z) & = \mr\,z + \Delta \,
\er^{-z/\eta}\,\cos\fetawp{x} \nonumber +
\frac{\xi^2}{\omega^2}\left(
\frac{2\,\mr\,\Delta^2}{\eta}\lp 1-\er^{-2z/\eta}\rp\right.\\
&\left.- \frac{\Delta^3}{2\,\eta^2}\lp \er^{-z/\eta}
-\er^{-3z/\eta}\rp\cos\fetawp{x} -
\frac{2\,\mr^2\,\Delta}{\eta}\,z\,\er^{-z/\eta}\,\cos\fetawp{x}\right)
+\BO{\eps^4}\;.
\end{align}
The above expansions have been carried out up to the first
nontrivial correction of the 0th-order approximation. Indeed, all
calculations must be pushed to $\BO{\eps^3}$ since an internal
$\xi$-layer is necessary to satisfy the boundary condition
(\ref{bcneu}) in $z=0$. This layer is of $\BO{\eps^3}$ because in
the Neumann case the boundary condition (\ref{bcneu}) concerns the
first derivative of $s$, instead of the degree of orientation
itself. We remark that the solutions
(\ref{s_sol_neu})-(\ref{theta_sol_neu}) are coherently ordered for
every fixed value of $\eta \neq 0$. However, they are not uniformly
ordered when $\eta \in (0,\bar{\eta}]$, namely we don't have a
uniform solution if $\eta$ is allowed to become of order $\xi$ or,
still worse, tend to zero. In other words, the above solutions
remain valid as $\eta \rightarrow 0^+$ if and only if $\xi =
\so{\eta}$. The main properties of the equilibrium configurations in
the mathematically appealing but physically uncommon case in which
$\eta$ is of the order of, or even smaller than, $\xi$ will be
presented elsewhere \cite{turphd}.

\subsubsection{Surface melting}

\begin{figure}[htp]
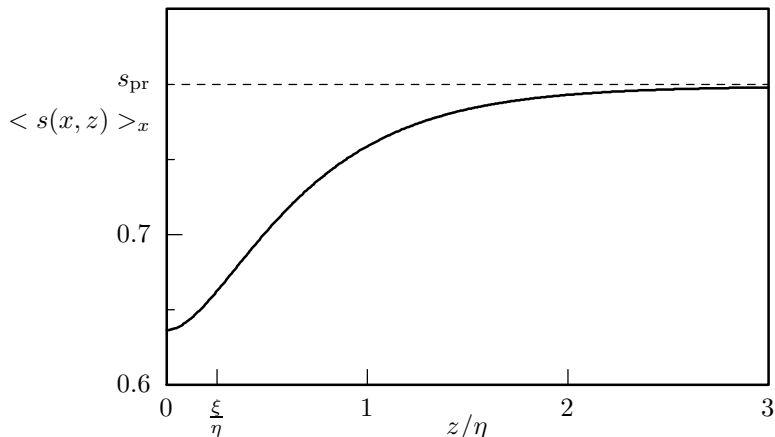

%
%
\begin{center}
\begin{texdraw}
\drawdim truecm \setgray 0
\linewd 0.03
\move (0 0) \lvec (8.0 0)
\lvec (8.0 5.0) \lvec (0 5.0) \lvec (0 0)
\linewd 0.02
\textref h:C v:T \htext (0.00 -.20) {0}
\move (0.667 0) \lvec (0.667 0.200)
\textref h:C v:T \htext (0.667 -.20) {$\frac{\xi}{\eta}$}
\move (2.667 0) \lvec (2.667 0.200)
\textref h:C v:T \htext (2.667 -.20) {1}
\move (5.333 0) \lvec (5.333 0.200)
\textref h:C v:T \htext (5.333 -.20) {2}
\textref h:C v:T \htext (8.00 -.20) {3}
\textref h:R v:C \htext (-.2 0.00) {0.6}
\move (0 1) \lvec (0.1 1)
\move (0 2) \lvec (0.2 2)
\textref h:R v:C \htext (-.2 2) {$0.7$}
\move (0 3) \lvec (0.1 3)
\lpatt (.1 .1)
\move (0 4) \lvec (8 4)
\textref h:R v:C \htext (-.2 4) {$\spr$}
\lpatt ()
\textref h:C v:T \htext (4 -.4) {$z/\eta$}
\textref h:R v:C \htext (-.2 3.5) {$<s(x,z)>_x$}
\linewd 0.03
\move(0.000 0.729) \lvec(0.016 0.730) \lvec(0.032 0.731) \lvec(0.048
0.733) \lvec(0.064 0.737) \lvec(0.080 0.741) \lvec(0.096 0.745)
\lvec(0.112 0.751) \lvec(0.128 0.757) \lvec(0.144 0.764) \lvec(0.160
0.772) \lvec(0.176 0.781) \lvec(0.192 0.790) \lvec(0.208 0.799)
\lvec(0.224 0.810) \lvec(0.240 0.820) \lvec(0.256 0.832) \lvec(0.272
0.843) \lvec(0.288 0.856) \lvec(0.304 0.869) \lvec(0.320 0.882)
\lvec(0.336 0.895) \lvec(0.352 0.910) \lvec(0.368 0.924) \lvec(0.384
0.939) \lvec(0.400 0.954) \lvec(0.416 0.970) \lvec(0.432 0.985)
\lvec(0.448 1.002) \lvec(0.464 1.018) \lvec(0.480 1.035) \lvec(0.496
1.052) \lvec(0.512 1.069) \lvec(0.528 1.086) \lvec(0.544 1.104)
\lvec(0.560 1.122) \lvec(0.576 1.140) \lvec(0.592 1.158) \lvec(0.608
1.176) \lvec(0.624 1.195) \lvec(0.640 1.214) \lvec(0.656 1.232)
\lvec(0.672 1.251) \lvec(0.688 1.270) \lvec(0.704 1.290) \lvec(0.720
1.309) \lvec(0.736 1.328) \lvec(0.752 1.348) \lvec(0.768 1.367)
\lvec(0.784 1.387) \lvec(0.800 1.406) \lvec(0.816 1.426) \lvec(0.832
1.445) \lvec(0.848 1.465) \lvec(0.864 1.485) \lvec(0.880 1.505)
\lvec(0.896 1.524) \lvec(0.912 1.544) \lvec(0.928 1.564) \lvec(0.944
1.584) \lvec(0.960 1.603) \lvec(0.976 1.623) \lvec(0.992 1.643)
\lvec(1.008 1.662) \lvec(1.024 1.682) \lvec(1.040 1.702) \lvec(1.056
1.721) \lvec(1.072 1.741) \lvec(1.088 1.760) \lvec(1.104 1.779)
\lvec(1.120 1.799) \lvec(1.136 1.818) \lvec(1.152 1.837) \lvec(1.168
1.856) \lvec(1.184 1.875) \lvec(1.200 1.894) \lvec(1.216 1.913)
\lvec(1.232 1.932) \lvec(1.248 1.951) \lvec(1.264 1.969) \lvec(1.280
1.988) \lvec(1.296 2.006) \lvec(1.312 2.025) \lvec(1.328 2.043)
\lvec(1.344 2.061) \lvec(1.360 2.079) \lvec(1.376 2.097) \lvec(1.392
2.115) \lvec(1.408 2.133) \lvec(1.424 2.151) \lvec(1.440 2.168)
\lvec(1.456 2.186) \lvec(1.472 2.203) \lvec(1.488 2.220) \lvec(1.504
2.238) \lvec(1.520 2.255) \lvec(1.536 2.272) \lvec(1.552 2.288)
\lvec(1.568 2.305) \lvec(1.584 2.322) \lvec(1.600 2.338) \lvec(1.616
2.354) \lvec(1.632 2.371) \lvec(1.648 2.387) \lvec(1.664 2.403)
\lvec(1.680 2.419) \lvec(1.696 2.434) \lvec(1.712 2.450) \lvec(1.728
2.465) \lvec(1.744 2.481) \lvec(1.760 2.496) \lvec(1.776 2.511)
\lvec(1.792 2.526) \lvec(1.808 2.541) \lvec(1.824 2.556) \lvec(1.840
2.571) \lvec(1.856 2.585) \lvec(1.872 2.600) \lvec(1.888 2.614)
\lvec(1.904 2.628) \lvec(1.920 2.642) \lvec(1.936 2.656) \lvec(1.952
2.670) \lvec(1.968 2.684) \lvec(1.984 2.697) \lvec(2.000 2.711)
\lvec(2.016 2.724) \lvec(2.032 2.737) \lvec(2.048 2.750) \lvec(2.064
2.763) \lvec(2.080 2.776) \lvec(2.096 2.789) \lvec(2.112 2.802)
\lvec(2.128 2.814) \lvec(2.144 2.827) \lvec(2.160 2.839) \lvec(2.176
2.851) \lvec(2.192 2.863) \lvec(2.208 2.875) \lvec(2.224 2.887)
\lvec(2.240 2.899) \lvec(2.256 2.911) \lvec(2.272 2.922) \lvec(2.288
2.934) \lvec(2.304 2.945) \lvec(2.320 2.956) \lvec(2.336 2.967)
\lvec(2.352 2.978) \lvec(2.368 2.989) \lvec(2.384 3.000) \lvec(2.400
3.010) \lvec(2.416 3.021) \lvec(2.432 3.031) \lvec(2.448 3.042)
\lvec(2.464 3.052) \lvec(2.480 3.062) \lvec(2.496 3.072) \lvec(2.512
3.082) \lvec(2.528 3.092) \lvec(2.544 3.102) \lvec(2.560 3.112)
\lvec(2.576 3.121) \lvec(2.592 3.131) \lvec(2.608 3.140) \lvec(2.624
3.149) \lvec(2.640 3.158) \lvec(2.656 3.167) \lvec(2.672 3.176)
\lvec(2.688 3.185) \lvec(2.704 3.194) \lvec(2.720 3.203) \lvec(2.736
3.211) \lvec(2.752 3.220) \lvec(2.768 3.228) \lvec(2.784 3.237)
\lvec(2.800 3.245) \lvec(2.816 3.253) \lvec(2.832 3.261) \lvec(2.848
3.269) \lvec(2.864 3.277) \lvec(2.880 3.285) \lvec(2.896 3.293)
\lvec(2.912 3.301) \lvec(2.928 3.308) \lvec(2.944 3.316) \lvec(2.960
3.323) \lvec(2.976 3.331) \lvec(2.992 3.338) \lvec(3.008 3.345)
\lvec(3.024 3.352) \lvec(3.040 3.359) \lvec(3.056 3.366) \lvec(3.072
3.373) \lvec(3.088 3.380) \lvec(3.104 3.387) \lvec(3.120 3.393)
\lvec(3.136 3.400) \lvec(3.152 3.407) \lvec(3.168 3.413) \lvec(3.184
3.419) \lvec(3.200 3.426) \lvec(3.216 3.432) \lvec(3.232 3.438)
\lvec(3.248 3.444) \lvec(3.264 3.450) \lvec(3.280 3.456) \lvec(3.296
3.462) \lvec(3.312 3.468) \lvec(3.328 3.474) \lvec(3.344 3.480)
\lvec(3.360 3.485) \lvec(3.376 3.491) \lvec(3.392 3.496) \lvec(3.408
3.502) \lvec(3.424 3.507) \lvec(3.440 3.513) \lvec(3.456 3.518)
\lvec(3.472 3.523) \lvec(3.488 3.528) \lvec(3.504 3.534) \lvec(3.520
3.539) \lvec(3.536 3.544) \lvec(3.552 3.549) \lvec(3.568 3.554)
\lvec(3.584 3.558) \lvec(3.600 3.563) \lvec(3.616 3.568) \lvec(3.632
3.573) \lvec(3.648 3.577) \lvec(3.664 3.582) \lvec(3.680 3.586)
\lvec(3.696 3.591) \lvec(3.712 3.595) \lvec(3.728 3.600) \lvec(3.744
3.604) \lvec(3.760 3.608) \lvec(3.776 3.612) \lvec(3.792 3.617)
\lvec(3.808 3.621) \lvec(3.824 3.625) \lvec(3.840 3.629) \lvec(3.856
3.633) \lvec(3.872 3.637) \lvec(3.888 3.641) \lvec(3.904 3.645)
\lvec(3.920 3.648) \lvec(3.936 3.652) \lvec(3.952 3.656) \lvec(3.968
3.660) \lvec(3.984 3.663) \lvec(4.000 3.667) \lvec(4.016 3.671)
\lvec(4.032 3.674) \lvec(4.048 3.678) \lvec(4.064 3.681) \lvec(4.080
3.684) \lvec(4.096 3.688) \lvec(4.112 3.691) \lvec(4.128 3.694)
\lvec(4.144 3.698) \lvec(4.160 3.701) \lvec(4.176 3.704) \lvec(4.192
3.707) \lvec(4.208 3.710) \lvec(4.224 3.713) \lvec(4.240 3.716)
\lvec(4.256 3.719) \lvec(4.272 3.722) \lvec(4.288 3.725) \lvec(4.304
3.728) \lvec(4.320 3.731) \lvec(4.336 3.734) \lvec(4.352 3.737)
\lvec(4.368 3.739) \lvec(4.384 3.742) \lvec(4.400 3.745) \lvec(4.416
3.748) \lvec(4.432 3.750) \lvec(4.448 3.753) \lvec(4.464 3.755)
\lvec(4.480 3.758) \lvec(4.496 3.760) \lvec(4.512 3.763) \lvec(4.528
3.765) \lvec(4.544 3.768) \lvec(4.560 3.770) \lvec(4.576 3.773)
\lvec(4.592 3.775) \lvec(4.608 3.777) \lvec(4.624 3.780) \lvec(4.640
3.782) \lvec(4.656 3.784) \lvec(4.672 3.786) \lvec(4.688 3.788)
\lvec(4.704 3.791) \lvec(4.720 3.793) \lvec(4.736 3.795) \lvec(4.752
3.797) \lvec(4.768 3.799) \lvec(4.784 3.801) \lvec(4.800 3.803)
\lvec(4.816 3.805) \lvec(4.832 3.807) \lvec(4.848 3.809) \lvec(4.864
3.811) \lvec(4.880 3.813) \lvec(4.896 3.815) \lvec(4.912 3.817)
\lvec(4.928 3.818) \lvec(4.944 3.820) \lvec(4.960 3.822) \lvec(4.976
3.824) \lvec(4.992 3.826) \lvec(5.008 3.827) \lvec(5.024 3.829)
\lvec(5.040 3.831) \lvec(5.056 3.832) \lvec(5.072 3.834) \lvec(5.088
3.836) \lvec(5.104 3.837) \lvec(5.120 3.839) \lvec(5.136 3.840)
\lvec(5.152 3.842) \lvec(5.168 3.844) \lvec(5.184 3.845) \lvec(5.200
3.847) \lvec(5.216 3.848) \lvec(5.232 3.850) \lvec(5.248 3.851)
\lvec(5.264 3.852) \lvec(5.280 3.854) \lvec(5.296 3.855) \lvec(5.312
3.857) \lvec(5.328 3.858) \lvec(5.344 3.859) \lvec(5.360 3.861)
\lvec(5.376 3.862) \lvec(5.392 3.863) \lvec(5.408 3.865) \lvec(5.424
3.866) \lvec(5.440 3.867) \lvec(5.456 3.868) \lvec(5.472 3.870)
\lvec(5.488 3.871) \lvec(5.504 3.872) \lvec(5.520 3.873) \lvec(5.536
3.874) \lvec(5.552 3.876) \lvec(5.568 3.877) \lvec(5.584 3.878)
\lvec(5.600 3.879) \lvec(5.616 3.880) \lvec(5.632 3.881) \lvec(5.648
3.882) \lvec(5.664 3.883) \lvec(5.680 3.884) \lvec(5.696 3.885)
\lvec(5.712 3.886) \lvec(5.728 3.887) \lvec(5.744 3.888) \lvec(5.760
3.889) \lvec(5.776 3.890) \lvec(5.792 3.891) \lvec(5.808 3.892)
\lvec(5.824 3.893) \lvec(5.840 3.894) \lvec(5.856 3.895) \lvec(5.872
3.896) \lvec(5.888 3.897) \lvec(5.904 3.898) \lvec(5.920 3.899)
\lvec(5.936 3.900) \lvec(5.952 3.901) \lvec(5.968 3.901) \lvec(5.984
3.902) \lvec(6.000 3.903) \lvec(6.016 3.904) \lvec(6.032 3.905)
\lvec(6.048 3.905) \lvec(6.064 3.906) \lvec(6.080 3.907) \lvec(6.096
3.908) \lvec(6.112 3.909) \lvec(6.128 3.909) \lvec(6.144 3.910)
\lvec(6.160 3.911) \lvec(6.176 3.912) \lvec(6.192 3.912) \lvec(6.208
3.913) \lvec(6.224 3.914) \lvec(6.240 3.914) \lvec(6.256 3.915)
\lvec(6.272 3.916) \lvec(6.288 3.916) \lvec(6.304 3.917) \lvec(6.320
3.918) \lvec(6.336 3.918) \lvec(6.352 3.919) \lvec(6.368 3.920)
\lvec(6.384 3.920) \lvec(6.400 3.921) \lvec(6.416 3.922) \lvec(6.432
3.922) \lvec(6.448 3.923) \lvec(6.464 3.923) \lvec(6.480 3.924)
\lvec(6.496 3.924) \lvec(6.512 3.925) \lvec(6.528 3.926) \lvec(6.544
3.926) \lvec(6.560 3.927) \lvec(6.576 3.927) \lvec(6.592 3.928)
\lvec(6.608 3.928) \lvec(6.624 3.929) \lvec(6.640 3.929) \lvec(6.656
3.930) \lvec(6.672 3.930) \lvec(6.688 3.931) \lvec(6.704 3.931)
\lvec(6.720 3.932) \lvec(6.736 3.932) \lvec(6.752 3.933) \lvec(6.768
3.933) \lvec(6.784 3.934) \lvec(6.800 3.934) \lvec(6.816 3.935)
\lvec(6.832 3.935) \lvec(6.848 3.936) \lvec(6.864 3.936) \lvec(6.880
3.936) \lvec(6.896 3.937) \lvec(6.912 3.937) \lvec(6.928 3.938)
\lvec(6.944 3.938) \lvec(6.960 3.938) \lvec(6.976 3.939) \lvec(6.992
3.939) \lvec(7.008 3.940) \lvec(7.024 3.940) \lvec(7.040 3.940)
\lvec(7.056 3.941) \lvec(7.072 3.941) \lvec(7.088 3.942) \lvec(7.104
3.942) \lvec(7.120 3.942) \lvec(7.136 3.943) \lvec(7.152 3.943)
\lvec(7.168 3.943) \lvec(7.184 3.944) \lvec(7.200 3.944) \lvec(7.216
3.944) \lvec(7.232 3.945) \lvec(7.248 3.945) \lvec(7.264 3.945)
\lvec(7.280 3.946) \lvec(7.296 3.946) \lvec(7.312 3.946) \lvec(7.328
3.947) \lvec(7.344 3.947) \lvec(7.360 3.947) \lvec(7.376 3.948)
\lvec(7.392 3.948) \lvec(7.408 3.948) \lvec(7.424 3.948) \lvec(7.440
3.949) \lvec(7.456 3.949) \lvec(7.472 3.949) \lvec(7.488 3.950)
\lvec(7.504 3.950) \lvec(7.520 3.950) \lvec(7.536 3.950) \lvec(7.552
3.951) \lvec(7.568 3.951) \lvec(7.584 3.951) \lvec(7.600 3.951)
\lvec(7.616 3.952) \lvec(7.632 3.952) \lvec(7.648 3.952) \lvec(7.664
3.952) \lvec(7.680 3.953) \lvec(7.696 3.953) \lvec(7.712 3.953)
\lvec(7.728 3.953) \lvec(7.744 3.953) \lvec(7.760 3.954) \lvec(7.776
3.954) \lvec(7.792 3.954) \lvec(7.808 3.954) \lvec(7.824 3.955)
\lvec(7.840 3.955) \lvec(7.856 3.955) \lvec(7.872 3.955) \lvec(7.888
3.955) \lvec(7.904 3.956) \lvec(7.920 3.956) \lvec(7.936 3.956)
\lvec(7.952 3.956) \lvec(7.968 3.956) \lvec(7.984 3.957) \lvec(8.000
3.957)
\end{texdraw}
\end{center}
%
%
\caption{Boundary layers in the mean degree of orientation
$<s(x,z)>_x$, when $\xi=0.25 \eta$, $\spr=0.8$, $\omega=0.6$,
$\mr=0.1/\eta$, and $\Delta=1.5$. The plot exhibits the presence of
\emph{two} boundary layers, the internal one being required by the
free boundary condition applied on $s$.} \label{s_sol_neu_gr1}
\end{figure}

We can highlight three different contributions in the degree of
orientation (\ref{s_sol_neu}). First, we notice a uniform decrease
in the degree of order, equal to $-\spr\mr^2 \xi^2/\omega^2$. This
disordering effect is triggered by the $\theta$-derivative $\mr$,
and was certainly to be expected. In fact, a glance to the free
energy functional (\ref{fb}) suffices to show that a reduction in
$s$ decreases the free energy whenever the gradient of the director
is not null. We then find two boundary layers. The former, of
thickness $\eta$ and $\BO{\eps^2}$, is a further reduction of the
degree of orientation due to the boundary roughness, that induces a
director variation in the $x$-direction. An internal boundary layer,
of thickness $\xi$ and order $\BO{\eps^3}$, is finally needed in
order to cancel the normal derivative of $s$ at the external
surface. If we take into account all the contributions, the mean
surface degree of orientation, defined as the $x$-average of
$s(x,0)$, turns out to be
\begin{equation}
 <s(x,0)>_{x}= \spr \qlp 1 - \frac{\mr^2\xi^2}{\omega^2}-
\frac{\Delta^2\xi^2}{\omega^2\eta^2}+
\frac{2\Delta^2\xi^3}{\sqrt{3}\,\omega^3\eta^3} \qrp
\;.\label{ssurf}
\end{equation}
Figure \ref{s_sol_neu_gr1} evidences the reported behaviour of the
mean degree of orientation as a function of the distance from the
surface.

\subsubsection{Effective surface angle}

\begin{figure}[htp]
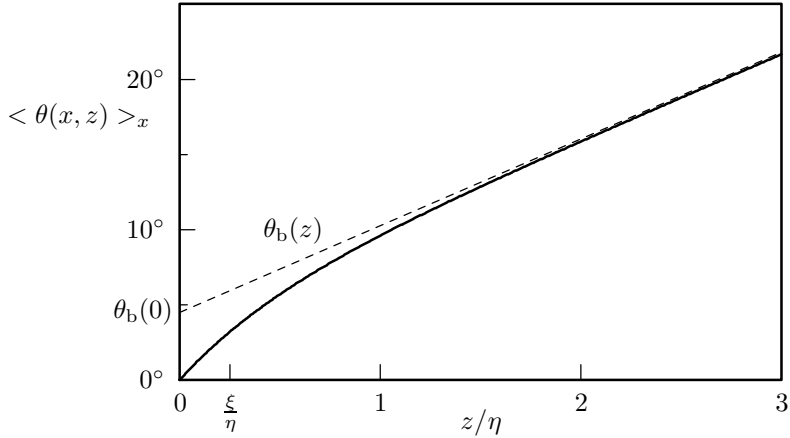

%
%
\begin{center}
\begin{texdraw}
\drawdim truecm \setgray 0
\linewd 0.03
\move (0 0) \lvec (8.0 0)
\lvec (8.0 5.0) \lvec (0 5.0) \lvec (0 0)
\linewd 0.02
\textref h:C v:T \htext (0.00 -.20) {0}
\move (0.667 0) \lvec (0.667 0.200)
\textref h:C v:T \htext (0.667 -.20) {$\frac{\xi}{\eta}$}
\move (2.667 0) \lvec (2.667 0.200)
\textref h:C v:T \htext (2.667 -.20) {1}
\move (5.333 0) \lvec (5.333 0.200)
\textref h:C v:T \htext (5.333 -.20) {2}
\textref h:C v:T \htext (8.00 -.20) {3}
\textref h:R v:C \htext (-.2 0) {$0^\circ$}
\move (0 1) \lvec (0.1 1)
\move (0 2) \lvec (0.2 2)
\textref h:R v:C \htext (-.2 2) {$10^\circ$}
\move (0 3) \lvec (0.1 3)
\move (0 4) \lvec (0.2 4)
\textref h:R v:C \htext (-.2 4) {$20^\circ$}
\lpatt (.1 .1)
\move(0.000 0.895247) \lvec(8.000 4.36)
\textref h:R v:C \htext (-.1 0.895247) {$\theta_{\rm b}(0)$}
\lpatt ()
\textref h:R v:B \htext (1.9 1.8) {$\theta_{\rm b}(z)$}
\textref h:C v:T \htext (4 -.4) {$z/\eta$}
\textref h:R v:C \htext (-.4 3.5) {$<\theta(x,z)>_x$}
\linewd 0.03
\move(0.000 0.000) \lvec(0.016 0.018) \lvec(0.032 0.035) \lvec(0.048
0.052) \lvec(0.064 0.069) \lvec(0.080 0.087) \lvec(0.096 0.103)
\lvec(0.112 0.120) \lvec(0.128 0.137) \lvec(0.144 0.154) \lvec(0.160
0.170) \lvec(0.176 0.186) \lvec(0.192 0.203) \lvec(0.208 0.219)
\lvec(0.224 0.235) \lvec(0.240 0.251) \lvec(0.256 0.266) \lvec(0.272
0.282) \lvec(0.288 0.298) \lvec(0.304 0.313) \lvec(0.320 0.329)
\lvec(0.336 0.344) \lvec(0.352 0.359) \lvec(0.368 0.374) \lvec(0.384
0.389) \lvec(0.400 0.404) \lvec(0.416 0.419) \lvec(0.432 0.433)
\lvec(0.448 0.448) \lvec(0.464 0.463) \lvec(0.480 0.477) \lvec(0.496
0.491) \lvec(0.512 0.506) \lvec(0.528 0.520) \lvec(0.544 0.534)
\lvec(0.560 0.548) \lvec(0.576 0.562) \lvec(0.592 0.575) \lvec(0.608
0.589) \lvec(0.624 0.603) \lvec(0.640 0.616) \lvec(0.656 0.630)
\lvec(0.672 0.643) \lvec(0.688 0.657) \lvec(0.704 0.670) \lvec(0.720
0.683) \lvec(0.736 0.696) \lvec(0.752 0.709) \lvec(0.768 0.722)
\lvec(0.784 0.735) \lvec(0.800 0.748) \lvec(0.816 0.760) \lvec(0.832
0.773) \lvec(0.848 0.786) \lvec(0.864 0.798) \lvec(0.880 0.811)
\lvec(0.896 0.823) \lvec(0.912 0.835) \lvec(0.928 0.848) \lvec(0.944
0.860) \lvec(0.960 0.872) \lvec(0.976 0.884) \lvec(0.992 0.896)
\lvec(1.008 0.908) \lvec(1.024 0.920) \lvec(1.040 0.932) \lvec(1.056
0.944) \lvec(1.072 0.955) \lvec(1.088 0.967) \lvec(1.104 0.979)
\lvec(1.120 0.990) \lvec(1.136 1.002) \lvec(1.152 1.013) \lvec(1.168
1.024) \lvec(1.184 1.036) \lvec(1.200 1.047) \lvec(1.216 1.058)
\lvec(1.232 1.069) \lvec(1.248 1.080) \lvec(1.264 1.092) \lvec(1.280
1.103) \lvec(1.296 1.114) \lvec(1.312 1.124) \lvec(1.328 1.135)
\lvec(1.344 1.146) \lvec(1.360 1.157) \lvec(1.376 1.168) \lvec(1.392
1.178) \lvec(1.408 1.189) \lvec(1.424 1.200) \lvec(1.440 1.210)
\lvec(1.456 1.221) \lvec(1.472 1.231) \lvec(1.488 1.241) \lvec(1.504
1.252) \lvec(1.520 1.262) \lvec(1.536 1.273) \lvec(1.552 1.283)
\lvec(1.568 1.293) \lvec(1.584 1.303) \lvec(1.600 1.313) \lvec(1.616
1.323) \lvec(1.632 1.333) \lvec(1.648 1.343) \lvec(1.664 1.353)
\lvec(1.680 1.363) \lvec(1.696 1.373) \lvec(1.712 1.383) \lvec(1.728
1.393) \lvec(1.744 1.403) \lvec(1.760 1.413) \lvec(1.776 1.422)
\lvec(1.792 1.432) \lvec(1.808 1.442) \lvec(1.824 1.451) \lvec(1.840
1.461) \lvec(1.856 1.470) \lvec(1.872 1.480) \lvec(1.888 1.489)
\lvec(1.904 1.499) \lvec(1.920 1.508) \lvec(1.936 1.518) \lvec(1.952
1.527) \lvec(1.968 1.536) \lvec(1.984 1.546) \lvec(2.000 1.555)
\lvec(2.016 1.564) \lvec(2.032 1.574) \lvec(2.048 1.583) \lvec(2.064
1.592) \lvec(2.080 1.601) \lvec(2.096 1.610) \lvec(2.112 1.619)
\lvec(2.128 1.628) \lvec(2.144 1.637) \lvec(2.160 1.646) \lvec(2.176
1.655) \lvec(2.192 1.664) \lvec(2.208 1.673) \lvec(2.224 1.682)
\lvec(2.240 1.691) \lvec(2.256 1.700) \lvec(2.272 1.709) \lvec(2.288
1.718) \lvec(2.304 1.726) \lvec(2.320 1.735) \lvec(2.336 1.744)
\lvec(2.352 1.753) \lvec(2.368 1.761) \lvec(2.384 1.770) \lvec(2.400
1.779) \lvec(2.416 1.787) \lvec(2.432 1.796) \lvec(2.448 1.805)
\lvec(2.464 1.813) \lvec(2.480 1.822) \lvec(2.496 1.830) \lvec(2.512
1.839) \lvec(2.528 1.847) \lvec(2.544 1.856) \lvec(2.560 1.864)
\lvec(2.576 1.873) \lvec(2.592 1.881) \lvec(2.608 1.890) \lvec(2.624
1.898) \lvec(2.640 1.906) \lvec(2.656 1.915) \lvec(2.672 1.923)
\lvec(2.688 1.931) \lvec(2.704 1.939) \lvec(2.720 1.948) \lvec(2.736
1.956) \lvec(2.752 1.964) \lvec(2.768 1.973) \lvec(2.784 1.981)
\lvec(2.800 1.989) \lvec(2.816 1.997) \lvec(2.832 2.005) \lvec(2.848
2.014) \lvec(2.864 2.022) \lvec(2.880 2.030) \lvec(2.896 2.038)
\lvec(2.912 2.046) \lvec(2.928 2.054) \lvec(2.944 2.062) \lvec(2.960
2.070) \lvec(2.976 2.078) \lvec(2.992 2.086) \lvec(3.008 2.094)
\lvec(3.024 2.102) \lvec(3.040 2.110) \lvec(3.056 2.118) \lvec(3.072
2.126) \lvec(3.088 2.134) \lvec(3.104 2.142) \lvec(3.120 2.150)
\lvec(3.136 2.158) \lvec(3.152 2.166) \lvec(3.168 2.174) \lvec(3.184
2.181) \lvec(3.200 2.189) \lvec(3.216 2.197) \lvec(3.232 2.205)
\lvec(3.248 2.213) \lvec(3.264 2.221) \lvec(3.280 2.228) \lvec(3.296
2.236) \lvec(3.312 2.244) \lvec(3.328 2.252) \lvec(3.344 2.259)
\lvec(3.360 2.267) \lvec(3.376 2.275) \lvec(3.392 2.283) \lvec(3.408
2.290) \lvec(3.424 2.298) \lvec(3.440 2.306) \lvec(3.456 2.314)
\lvec(3.472 2.321) \lvec(3.488 2.329) \lvec(3.504 2.336) \lvec(3.520
2.344) \lvec(3.536 2.352) \lvec(3.552 2.359) \lvec(3.568 2.367)
\lvec(3.584 2.375) \lvec(3.600 2.382) \lvec(3.616 2.390) \lvec(3.632
2.397) \lvec(3.648 2.405) \lvec(3.664 2.413) \lvec(3.680 2.420)
\lvec(3.696 2.428) \lvec(3.712 2.435) \lvec(3.728 2.443) \lvec(3.744
2.450) \lvec(3.760 2.458) \lvec(3.776 2.465) \lvec(3.792 2.473)
\lvec(3.808 2.480) \lvec(3.824 2.488) \lvec(3.840 2.495) \lvec(3.856
2.503) \lvec(3.872 2.510) \lvec(3.888 2.518) \lvec(3.904 2.525)
\lvec(3.920 2.533) \lvec(3.936 2.540) \lvec(3.952 2.547) \lvec(3.968
2.555) \lvec(3.984 2.562) \lvec(4.000 2.570) \lvec(4.016 2.577)
\lvec(4.032 2.585) \lvec(4.048 2.592) \lvec(4.064 2.599) \lvec(4.080
2.607) \lvec(4.096 2.614) \lvec(4.112 2.621) \lvec(4.128 2.629)
\lvec(4.144 2.636) \lvec(4.160 2.644) \lvec(4.176 2.651) \lvec(4.192
2.658) \lvec(4.208 2.666) \lvec(4.224 2.673) \lvec(4.240 2.680)
\lvec(4.256 2.687) \lvec(4.272 2.695) \lvec(4.288 2.702) \lvec(4.304
2.709) \lvec(4.320 2.717) \lvec(4.336 2.724) \lvec(4.352 2.731)
\lvec(4.368 2.739) \lvec(4.384 2.746) \lvec(4.400 2.753) \lvec(4.416
2.760) \lvec(4.432 2.768) \lvec(4.448 2.775) \lvec(4.464 2.782)
\lvec(4.480 2.789) \lvec(4.496 2.797) \lvec(4.512 2.804) \lvec(4.528
2.811) \lvec(4.544 2.818) \lvec(4.560 2.826) \lvec(4.576 2.833)
\lvec(4.592 2.840) \lvec(4.608 2.847) \lvec(4.624 2.855) \lvec(4.640
2.862) \lvec(4.656 2.869) \lvec(4.672 2.876) \lvec(4.688 2.883)
\lvec(4.704 2.891) \lvec(4.720 2.898) \lvec(4.736 2.905) \lvec(4.752
2.912) \lvec(4.768 2.919) \lvec(4.784 2.926) \lvec(4.800 2.934)
\lvec(4.816 2.941) \lvec(4.832 2.948) \lvec(4.848 2.955) \lvec(4.864
2.962) \lvec(4.880 2.969) \lvec(4.896 2.977) \lvec(4.912 2.984)
\lvec(4.928 2.991) \lvec(4.944 2.998) \lvec(4.960 3.005) \lvec(4.976
3.012) \lvec(4.992 3.019) \lvec(5.008 3.027) \lvec(5.024 3.034)
\lvec(5.040 3.041) \lvec(5.056 3.048) \lvec(5.072 3.055) \lvec(5.088
3.062) \lvec(5.104 3.069) \lvec(5.120 3.076) \lvec(5.136 3.084)
\lvec(5.152 3.091) \lvec(5.168 3.098) \lvec(5.184 3.105) \lvec(5.200
3.112) \lvec(5.216 3.119) \lvec(5.232 3.126) \lvec(5.248 3.133)
\lvec(5.264 3.140) \lvec(5.280 3.147) \lvec(5.296 3.154) \lvec(5.312
3.161) \lvec(5.328 3.169) \lvec(5.344 3.176) \lvec(5.360 3.183)
\lvec(5.376 3.190) \lvec(5.392 3.197) \lvec(5.408 3.204) \lvec(5.424
3.211) \lvec(5.440 3.218) \lvec(5.456 3.225) \lvec(5.472 3.232)
\lvec(5.488 3.239) \lvec(5.504 3.246) \lvec(5.520 3.253) \lvec(5.536
3.260) \lvec(5.552 3.267) \lvec(5.568 3.274) \lvec(5.584 3.281)
\lvec(5.600 3.288) \lvec(5.616 3.296) \lvec(5.632 3.303) \lvec(5.648
3.310) \lvec(5.664 3.317) \lvec(5.680 3.324) \lvec(5.696 3.331)
\lvec(5.712 3.338) \lvec(5.728 3.345) \lvec(5.744 3.352) \lvec(5.760
3.359) \lvec(5.776 3.366) \lvec(5.792 3.373) \lvec(5.808 3.380)
\lvec(5.824 3.387) \lvec(5.840 3.394) \lvec(5.856 3.401) \lvec(5.872
3.408) \lvec(5.888 3.415) \lvec(5.904 3.422) \lvec(5.920 3.429)
\lvec(5.936 3.436) \lvec(5.952 3.443) \lvec(5.968 3.450) \lvec(5.984
3.457) \lvec(6.000 3.464) \lvec(6.016 3.471) \lvec(6.032 3.478)
\lvec(6.048 3.485) \lvec(6.064 3.492) \lvec(6.080 3.499) \lvec(6.096
3.506) \lvec(6.112 3.513) \lvec(6.128 3.520) \lvec(6.144 3.527)
\lvec(6.160 3.534) \lvec(6.176 3.541) \lvec(6.192 3.548) \lvec(6.208
3.555) \lvec(6.224 3.562) \lvec(6.240 3.569) \lvec(6.256 3.576)
\lvec(6.272 3.583) \lvec(6.288 3.590) \lvec(6.304 3.596) \lvec(6.320
3.603) \lvec(6.336 3.610) \lvec(6.352 3.617) \lvec(6.368 3.624)
\lvec(6.384 3.631) \lvec(6.400 3.638) \lvec(6.416 3.645) \lvec(6.432
3.652) \lvec(6.448 3.659) \lvec(6.464 3.666) \lvec(6.480 3.673)
\lvec(6.496 3.680) \lvec(6.512 3.687) \lvec(6.528 3.694) \lvec(6.544
3.701) \lvec(6.560 3.708) \lvec(6.576 3.715) \lvec(6.592 3.722)
\lvec(6.608 3.729) \lvec(6.624 3.736) \lvec(6.640 3.743) \lvec(6.656
3.750) \lvec(6.672 3.757) \lvec(6.688 3.764) \lvec(6.704 3.770)
\lvec(6.720 3.777) \lvec(6.736 3.784) \lvec(6.752 3.791) \lvec(6.768
3.798) \lvec(6.784 3.805) \lvec(6.800 3.812) \lvec(6.816 3.819)
\lvec(6.832 3.826) \lvec(6.848 3.833) \lvec(6.864 3.840) \lvec(6.880
3.847) \lvec(6.896 3.854) \lvec(6.912 3.861) \lvec(6.928 3.868)
\lvec(6.944 3.875) \lvec(6.960 3.882) \lvec(6.976 3.888) \lvec(6.992
3.895) \lvec(7.008 3.902) \lvec(7.024 3.909) \lvec(7.040 3.916)
\lvec(7.056 3.923) \lvec(7.072 3.930) \lvec(7.088 3.937) \lvec(7.104
3.944) \lvec(7.120 3.951) \lvec(7.136 3.958) \lvec(7.152 3.965)
\lvec(7.168 3.972) \lvec(7.184 3.979) \lvec(7.200 3.985) \lvec(7.216
3.992) \lvec(7.232 3.999) \lvec(7.248 4.006) \lvec(7.264 4.013)
\lvec(7.280 4.020) \lvec(7.296 4.027) \lvec(7.312 4.034) \lvec(7.328
4.041) \lvec(7.344 4.048) \lvec(7.360 4.055) \lvec(7.376 4.062)
\lvec(7.392 4.069) \lvec(7.408 4.075) \lvec(7.424 4.082) \lvec(7.440
4.089) \lvec(7.456 4.096) \lvec(7.472 4.103) \lvec(7.488 4.110)
\lvec(7.504 4.117) \lvec(7.520 4.124) \lvec(7.536 4.131) \lvec(7.552
4.138) \lvec(7.568 4.145) \lvec(7.584 4.152) \lvec(7.600 4.158)
\lvec(7.616 4.165) \lvec(7.632 4.172) \lvec(7.648 4.179) \lvec(7.664
4.186) \lvec(7.680 4.193) \lvec(7.696 4.200) \lvec(7.712 4.207)
\lvec(7.728 4.214) \lvec(7.744 4.221) \lvec(7.760 4.227) \lvec(7.776
4.234) \lvec(7.792 4.241) \lvec(7.808 4.248) \lvec(7.824 4.255)
\lvec(7.840 4.262) \lvec(7.856 4.269) \lvec(7.872 4.276) \lvec(7.888
4.283) \lvec(7.904 4.290) \lvec(7.920 4.297) \lvec(7.936 4.303)
\lvec(7.952 4.310) \lvec(7.968 4.317) \lvec(7.984 4.324) \lvec(8.000
4.331)
\end{texdraw}
\end{center}
%
%
\caption{Boundary layer in the mean tilt angle $<\theta(x,z)>_x$,
when $\xi=0.25 \eta$, $\spr=0.8$, $\omega=0.6$, $\mr=0.1/\eta$, and
$\Delta=1.5$. The dashed line corresponds to the asymptotic, linear
approximation $\theta_{\rm b}(z)$.} \label{theta_sol_neu_gr1}
\end{figure}

The tilt angle $\theta$ exhibits a boundary-layer structure as well.
Equation (\ref{theta_sol_neu}) shows that such layer is of
$\BO{\eps^2}$ and thickness $\eta$. It gives rise to an interesting
effective misalignment of the surface director. Indeed, if we allow
$z\gg\eta$ in (\ref{theta_sol_neu}) we find that
\begin{equation}
\label{theta_bulk} \theta(x,z)\approx \theta_{\rm b}(z)=
\frac{2\,\mr\,\xi^2\Delta^2}{\eta\omega^2}+\mr\,z \qquad {\rm
as}\quad z \gg\eta\;.
\end{equation}
The asymptotic approximation (\ref{theta_bulk}) shows that an
experimental observation, performed sufficiently far from the
external plate (with respect to the microscopic scale $\eta$) would
detect an \emph{effective\/} tilt angle $\theta_{\rm b}\/$, whose
value at the plate is different from zero, since
\begin{equation}
\theta_{\rm b}(0)=\frac{2\,\mr\,\xi^2\Delta^2}{\eta\omega^2}\;.
\label{thetapp}
\end{equation}
Thus, a coarse observation of the nematic configuration measures a
surface tilt angle slightly different from the homeotropic
prescription $\theta_{\rm surf}=0$. Figure \ref{theta_sol_neu_gr1}
evidences this effect. In the next section we will analyze in more
detail the result (\ref{thetapp}). Then we will show how it matches
the predictions of an effective weak anchoring potential. We remark
that the tilt angle does not exhibit any further boundary layer at
the smaller scale $\xi$.

\subsection{Fixed surface degree of orientation}\label{subsec:dir}

The perturbative analysis of the differential equations
(\ref{EL0102}), with the Dirichlet boundary conditions
(\ref{bcdir}), would be unnecessarily entangled because of the
non-linearity of the thermodynamic potential (\ref{therpot}). In
fact, in this case only implicit solutions for $s_0(x,z,Z)$ can be
gathered. In order to pursue our analysis, and still catch the
essential features of the solutions, we replace the function
$\sigma$ in (\ref{EL0102}) by its linear approximation
$\sigma_1(s)=\omega^2(s-\spr)$. This is tantamount to replacing the
Landau-de Gennes potential in (\ref{potsn}) by a tangent quadratic
well, still centered in $\spr$. Such approximation is certainly
well-justified deep in the nematic phase, when the isotropic state
$s=0$ becomes unstable, and the second well of the Landau-de Gennes
potential can be neglected.

The asymptotic properties of the solutions in this case depend
critically on the value $\tilde s$ forced on the surface. If $\tilde
s\neq \spr$, the boundary layer induced by the Dirichlet condition
dominates over the roughness effect. Indeed, the leading asymptotic
solutions are given by
\begin{align}
s(x,z)&= \spr - \lp \spr - \tilde s\rp \, \er^{-\sqrt{3}\,\omega
z/\xi}\nonumber \\
&\phantom{\null=\spr }- \sqrt{3}\,\lp \spr - \tilde s\rp \,
\frac{\xi}{\omega} \, \er^{-\sqrt{3}\,\omega z/\xi}\qlp
\frac{\Delta^2}{4 \eta}\, \lp 1-\er^{-2z/\eta}\rp +
\frac{3}{2}\,\mr^2\,z\right. \nonumber \\
&\left. \phantom{=\spr } -3 \mr \Delta \lp 1-\er^{-z/\eta} \rp \cos
\fetawp{x} + \frac{\Delta^2}{2 \eta} \lp 1-\er^{-2z/\eta} \rp
\cos \fetawp{2x}\qrp+\BO{\eps^2} \label{sdir1}\\
\theta(x,z)&= \mr\,z + \Delta \er^{-z/\eta} \cos\fetawp{x}\nonumber \\
&\phantom{\null=\mr\, z}+ \frac{\xi}{\sqrt{3}\,\omega} \qlp
h\lp\frac{z}{\xi}\rp - h(0) \qrp \lp \mr - \frac{\Delta}{\eta}
\er^{-z/\eta} \cos\fetawp{x} \rp +\BO{\eps^2}\,, \label{ttdir1}
\end{align}
where
\begin{equation}
h(\zeta)=\log \qlp\spr - \lp\spr - \tilde s\rp
\er^{-\sqrt{3}\,\omega \, \zeta}\qrp - \frac{\lp \spr - \tilde s\rp
\, \er^{-\sqrt{3}\,\omega \, \zeta}}{\spr - \lp \spr - \tilde s\rp
\er^{-\sqrt{3}\,\omega \, \zeta}}
\end{equation}
determines the tilt angle variation within the boundary-layer. The
bulk-asymptotic tilt angle is then given by
\begin{equation}
\label{theta_bulk_dir} \theta(x,z)\approx \theta_{\rm b}(z)=
\frac{\mr\,\xi}{\sqrt{3}\,\omega}\left(\log \frac{\spr}{\tilde s} +
\frac{\spr-\tilde s}{\tilde s} \right) +\mr\,z \qquad {\rm as}\quad
z \gg\eta\;.
\end{equation}
We remark that, when $\tilde s\neq \spr$, the leading contribution
to $\theta_{\rm b}(0)$ is independent of $\Delta$, and thus does not
depend on the surface roughness. Furthermore, the effective surface
tilt angle depends linearly on $\xi$, which makes it significantly
larger than the prediction (\ref{thetapp}), derived with
Neumann-like boundary conditions on $s$, which possesses an extra
$\xi/\eta$ (small) factor. Finally, we remark the fact that
$\theta_{\rm b}(0)$ shares the sign of $\mr$ if and only if $\tilde
s<\spr$. We will return below on the physical origin and
implications of this result.

When the induced degree of orientation $\tilde s$ does exactly
coincide with $\spr$, all calculations simplify since
$h(\zeta)\equiv\log \spr $, and all first order correction in
(\ref{ttdir1}) vanish. We therefore push our perturbation analysis,
and obtain
\begin{align}
s(x,z)&= \spr - \frac{\spr \xi^2}{\omega^2} \qlp \mr^2 +
\frac{\Delta^2}{\eta^2} \er^{-2z/\eta} -
\frac{2\mr\,\Delta}{\eta}\er^{-z/\eta}\cos\fetawp{x}\right.
\nonumber\\
&\left. \phantom{=\spr }-\expeps\,\lp \mr^2 +
\frac{\Delta^2}{\eta^2} - \frac{2\mr\,\Delta}{\eta}\cos\fetawp{x}
\rp \qrp +\BO{\eps^3}
\label{sdir2}\\
\theta(x,z) & = \mr\,z + \Delta \, \er^{-z/\eta}\,\cos\fetawp{x}
\nonumber + \frac{\xi^2}{\omega^2}\left(
\frac{2\,\mr\,\Delta^2}{\eta}\lp 1-\er^{-2z/\eta}\rp\right.\\
&\left.- \frac{\Delta^3}{2\,\eta^2}\lp \er^{-z/\eta}
-\er^{-3z/\eta}\rp\cos\fetawp{x} -
\frac{2\,\mr^2\,\Delta}{\eta}\,z\,\er^{-z/\eta}\,\cos\fetawp{x}\right)
+\BO{\eps^3}\;.\label{thetadir2}
\end{align}
Equation (\ref{thetadir2}) allows to compute the asymptotic tilt
angle $\theta_{\rm b}$, when $\tilde s=\spr$. In fact, once we drop
all exponentially-decaying terms in (\ref{thetadir2}), we arrive at
the interesting result that $\theta_{\rm b}(z)$ does exactly
coincide with (\ref{theta_bulk}), that is with the expression we
derived with a Neumann-like boundary condition on the degree of
orientation. In fact, the complete expression (\ref{thetadir2}) for
the tilt angle $\theta(x,z)$ coincides with (\ref{theta_sol_neu}) up
to $\BO{\eps^3}$. Thus, any observation on the tilt angle is not
able to distinguish among a free and a fixed boundary condition on
the degree of orientation, as long as the imposed value $\tilde s$
coincides with the preferred value $\spr$. This similarity between
the Neumann and Dirichlet cases can be pursued further. Indeed, we
can determine the $\BO{\eps^2}$-contributions in
(\ref{sdir1})-(\ref{ttdir1}) also when $\tilde s\neq \spr$. If we
then use them to compute the $\BO{\eps^2}$-correction to the
asymptotic tilt angle (\ref{theta_bulk_dir}), we arrive at the
following expression, valid at $\BO{\eps^2}$ for any value of
$\tilde s$:
\begin{equation}
\label{thbdir} \theta(x,z)\approx \theta_{\rm b}(z)=\left[
\frac{\mr\,\xi}{\sqrt{3}\,\omega}\left(\log \frac{\spr}{\tilde s} +
\frac{\spr-\tilde s}{\tilde s} \right)
+\frac{2\,\mr\,\xi^2\Delta^2}{\eta\omega^2}\right]+\mr\,z \quad {\rm
as}\ z \gg\eta,
\end{equation}
that yields
\begin{equation}
\label{ttdir0} \theta_{\rm b}(0)=
\frac{\mr\,\xi}{\sqrt{3}\,\omega}\left(\log \frac{\spr}{\tilde s} +
\frac{\spr-\tilde s}{\tilde s} \right)
+\frac{2\,\mr\,\xi^2\Delta^2}{\eta\omega^2}\;.
\end{equation}
The $\BO{\eps^2}$-contribution to the effective surface angle
$\theta_{\rm b}(0)$ is thus fully a roughness effect, and does not
depend at all on the type of boundary conditions imposed on $s$. On
the other hand, equation (\ref{ttdir0}) confirms that the effective
surface angle possesses also an $\BO{\eps}$-term when Dirichlet
conditions are imposed on the degree of orientation, and $\tilde
s\neq \spr$.

\begin{figure}[htp]
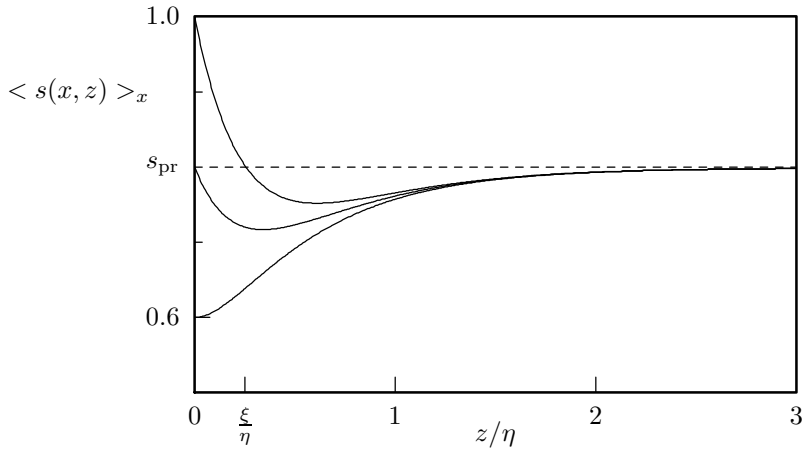

%
%
\begin{center}
\begin{texdraw}
\drawdim truecm \setgray 0
\linewd 0.03
\move (0 0) \lvec (8.0 0)
\lvec (8.0 5.0) \lvec (0 5.0) \lvec (0 0)
\linewd 0.02
\textref h:C v:T \htext (0.00 -.20) {0}
\move (0.667 0) \lvec (0.667 0.200)
\textref h:C v:T \htext (0.667 -.20) {$\frac{\xi}{\eta}$}
\move (2.667 0) \lvec (2.667 0.200)
\textref h:C v:T \htext (2.667 -.20) {1}
\move (5.333 0) \lvec (5.333 0.200)
\textref h:C v:T \htext (5.333 -.20) {2}
\textref h:C v:T \htext (8.00 -.20) {3}
\move (0 1.000) \lvec (0.20 1.000)
\textref h:R v:C \htext (-.2 1.00) {0.6}
\move (0 2.000) \lvec (0.10 2.000)
\lpatt (.1 .1)
\move (0 3.000) \lvec (8 3.000)
\lpatt ()
\textref h:R v:C \htext (-.2 3.00) {$\spr$}
\move (0 4.000) \lvec (0.10 4.000)
\textref h:R v:C \htext (-.2 5.00) {1.0}
\textref h:C v:T \htext (4 -.4) {$z/\eta$}
\textref h:R v:C \htext (-.6 4) {$<s(x,z)>_x$}
\move(0.000 5.000) \lvec(0.016 4.920) \lvec(0.032 4.843) \lvec(0.048
4.767) \lvec(0.064 4.693) \lvec(0.080 4.622) \lvec(0.096 4.552)
\lvec(0.112 4.484) \lvec(0.128 4.418) \lvec(0.144 4.354) \lvec(0.160
4.292) \lvec(0.176 4.231) \lvec(0.192 4.172) \lvec(0.208 4.114)
\lvec(0.224 4.058) \lvec(0.240 4.004) \lvec(0.256 3.951) \lvec(0.272
3.900) \lvec(0.288 3.851) \lvec(0.304 3.802) \lvec(0.320 3.755)
\lvec(0.336 3.710) \lvec(0.352 3.665) \lvec(0.368 3.622) \lvec(0.384
3.581) \lvec(0.400 3.540) \lvec(0.416 3.501) \lvec(0.432 3.463)
\lvec(0.448 3.426) \lvec(0.464 3.390) \lvec(0.480 3.356) \lvec(0.496
3.322) \lvec(0.512 3.289) \lvec(0.528 3.258) \lvec(0.544 3.227)
\lvec(0.560 3.198) \lvec(0.576 3.169) \lvec(0.592 3.141) \lvec(0.608
3.114) \lvec(0.624 3.089) \lvec(0.640 3.063) \lvec(0.656 3.039)
\lvec(0.672 3.016) \lvec(0.688 2.993) \lvec(0.704 2.971) \lvec(0.720
2.950) \lvec(0.736 2.929) \lvec(0.752 2.910) \lvec(0.768 2.891)
\lvec(0.784 2.872) \lvec(0.800 2.855) \lvec(0.816 2.838) \lvec(0.832
2.821) \lvec(0.848 2.806) \lvec(0.864 2.790) \lvec(0.880 2.776)
\lvec(0.896 2.762) \lvec(0.912 2.748) \lvec(0.928 2.735) \lvec(0.944
2.723) \lvec(0.960 2.711) \lvec(0.976 2.699) \lvec(0.992 2.688)
\lvec(1.008 2.678) \lvec(1.024 2.668) \lvec(1.040 2.658) \lvec(1.056
2.649) \lvec(1.072 2.640) \lvec(1.088 2.631) \lvec(1.104 2.623)
\lvec(1.120 2.616) \lvec(1.136 2.609) \lvec(1.152 2.602) \lvec(1.168
2.595) \lvec(1.184 2.589) \lvec(1.200 2.583) \lvec(1.216 2.577)
\lvec(1.232 2.572) \lvec(1.248 2.567) \lvec(1.264 2.562) \lvec(1.280
2.558) \lvec(1.296 2.554) \lvec(1.312 2.550) \lvec(1.328 2.546)
\lvec(1.344 2.543) \lvec(1.360 2.540) \lvec(1.376 2.537) \lvec(1.392
2.534) \lvec(1.408 2.532) \lvec(1.424 2.529) \lvec(1.440 2.527)
\lvec(1.456 2.526) \lvec(1.472 2.524) \lvec(1.488 2.523) \lvec(1.504
2.521) \lvec(1.520 2.520) \lvec(1.536 2.519) \lvec(1.552 2.519)
\lvec(1.568 2.518) \lvec(1.584 2.517) \lvec(1.600 2.517) \lvec(1.616
2.517) \lvec(1.632 2.517) \lvec(1.648 2.517) \lvec(1.664 2.517)
\lvec(1.680 2.518) \lvec(1.696 2.518) \lvec(1.712 2.519) \lvec(1.728
2.519) \lvec(1.744 2.520) \lvec(1.760 2.521) \lvec(1.776 2.522)
\lvec(1.792 2.523) \lvec(1.808 2.524) \lvec(1.824 2.526) \lvec(1.840
2.527) \lvec(1.856 2.529) \lvec(1.872 2.530) \lvec(1.888 2.532)
\lvec(1.904 2.534) \lvec(1.920 2.535) \lvec(1.936 2.537) \lvec(1.952
2.539) \lvec(1.968 2.541) \lvec(1.984 2.543) \lvec(2.000 2.545)
\lvec(2.016 2.547) \lvec(2.032 2.549) \lvec(2.048 2.551) \lvec(2.064
2.554) \lvec(2.080 2.556) \lvec(2.096 2.558) \lvec(2.112 2.561)
\lvec(2.128 2.563) \lvec(2.144 2.566) \lvec(2.160 2.568) \lvec(2.176
2.570) \lvec(2.192 2.573) \lvec(2.208 2.576) \lvec(2.224 2.578)
\lvec(2.240 2.581) \lvec(2.256 2.584) \lvec(2.272 2.586) \lvec(2.288
2.589) \lvec(2.304 2.592) \lvec(2.320 2.594) \lvec(2.336 2.597)
\lvec(2.352 2.600) \lvec(2.368 2.602) \lvec(2.384 2.605) \lvec(2.400
2.608) \lvec(2.416 2.611) \lvec(2.432 2.614) \lvec(2.448 2.616)
\lvec(2.464 2.619) \lvec(2.480 2.622) \lvec(2.496 2.625) \lvec(2.512
2.628) \lvec(2.528 2.631) \lvec(2.544 2.633) \lvec(2.560 2.636)
\lvec(2.576 2.639) \lvec(2.592 2.642) \lvec(2.608 2.645) \lvec(2.624
2.648) \lvec(2.640 2.650) \lvec(2.656 2.653) \lvec(2.672 2.656)
\lvec(2.688 2.659) \lvec(2.704 2.661) \lvec(2.720 2.664) \lvec(2.736
2.667) \lvec(2.752 2.670) \lvec(2.768 2.673) \lvec(2.784 2.675)
\lvec(2.800 2.678) \lvec(2.816 2.681) \lvec(2.832 2.684) \lvec(2.848
2.686) \lvec(2.864 2.689) \lvec(2.880 2.692) \lvec(2.896 2.694)
\lvec(2.912 2.697) \lvec(2.928 2.700) \lvec(2.944 2.702) \lvec(2.960
2.705) \lvec(2.976 2.708) \lvec(2.992 2.710) \lvec(3.008 2.713)
\lvec(3.024 2.715) \lvec(3.040 2.718) \lvec(3.056 2.721) \lvec(3.072
2.723) \lvec(3.088 2.726) \lvec(3.104 2.728) \lvec(3.120 2.731)
\lvec(3.136 2.733) \lvec(3.152 2.736) \lvec(3.168 2.738) \lvec(3.184
2.740) \lvec(3.200 2.743) \lvec(3.216 2.745) \lvec(3.232 2.748)
\lvec(3.248 2.750) \lvec(3.264 2.752) \lvec(3.280 2.755) \lvec(3.296
2.757) \lvec(3.312 2.759) \lvec(3.328 2.762) \lvec(3.344 2.764)
\lvec(3.360 2.766) \lvec(3.376 2.768) \lvec(3.392 2.771) \lvec(3.408
2.773) \lvec(3.424 2.775) \lvec(3.440 2.777) \lvec(3.456 2.779)
\lvec(3.472 2.781) \lvec(3.488 2.783) \lvec(3.504 2.786) \lvec(3.520
2.788) \lvec(3.536 2.790) \lvec(3.552 2.792) \lvec(3.568 2.794)
\lvec(3.584 2.796) \lvec(3.600 2.798) \lvec(3.616 2.800) \lvec(3.632
2.802) \lvec(3.648 2.804) \lvec(3.664 2.806) \lvec(3.680 2.807)
\lvec(3.696 2.809) \lvec(3.712 2.811) \lvec(3.728 2.813) \lvec(3.744
2.815) \lvec(3.760 2.817) \lvec(3.776 2.818) \lvec(3.792 2.820)
\lvec(3.808 2.822) \lvec(3.824 2.824) \lvec(3.840 2.826) \lvec(3.856
2.827) \lvec(3.872 2.829) \lvec(3.888 2.831) \lvec(3.904 2.832)
\lvec(3.920 2.834) \lvec(3.936 2.836) \lvec(3.952 2.837) \lvec(3.968
2.839) \lvec(3.984 2.841) \lvec(4.000 2.842) \lvec(4.016 2.844)
\lvec(4.032 2.845) \lvec(4.048 2.847) \lvec(4.064 2.848) \lvec(4.080
2.850) \lvec(4.096 2.851) \lvec(4.112 2.853) \lvec(4.128 2.854)
\lvec(4.144 2.856) \lvec(4.160 2.857) \lvec(4.176 2.859) \lvec(4.192
2.860) \lvec(4.208 2.861) \lvec(4.224 2.863) \lvec(4.240 2.864)
\lvec(4.256 2.865) \lvec(4.272 2.867) \lvec(4.288 2.868) \lvec(4.304
2.869) \lvec(4.320 2.871) \lvec(4.336 2.872) \lvec(4.352 2.873)
\lvec(4.368 2.875) \lvec(4.384 2.876) \lvec(4.400 2.877) \lvec(4.416
2.878) \lvec(4.432 2.879) \lvec(4.448 2.881) \lvec(4.464 2.882)
\lvec(4.480 2.883) \lvec(4.496 2.884) \lvec(4.512 2.885) \lvec(4.528
2.886) \lvec(4.544 2.888) \lvec(4.560 2.889) \lvec(4.576 2.890)
\lvec(4.592 2.891) \lvec(4.608 2.892) \lvec(4.624 2.893) \lvec(4.640
2.894) \lvec(4.656 2.895) \lvec(4.672 2.896) \lvec(4.688 2.897)
\lvec(4.704 2.898) \lvec(4.720 2.899) \lvec(4.736 2.900) \lvec(4.752
2.901) \lvec(4.768 2.902) \lvec(4.784 2.903) \lvec(4.800 2.904)
\lvec(4.816 2.905) \lvec(4.832 2.906) \lvec(4.848 2.907) \lvec(4.864
2.908) \lvec(4.880 2.909) \lvec(4.896 2.910) \lvec(4.912 2.910)
\lvec(4.928 2.911) \lvec(4.944 2.912) \lvec(4.960 2.913) \lvec(4.976
2.914) \lvec(4.992 2.915) \lvec(5.008 2.915) \lvec(5.024 2.916)
\lvec(5.040 2.917) \lvec(5.056 2.918) \lvec(5.072 2.919) \lvec(5.088
2.919) \lvec(5.104 2.920) \lvec(5.120 2.921) \lvec(5.136 2.922)
\lvec(5.152 2.922) \lvec(5.168 2.923) \lvec(5.184 2.924) \lvec(5.200
2.925) \lvec(5.216 2.925) \lvec(5.232 2.926) \lvec(5.248 2.927)
\lvec(5.264 2.928) \lvec(5.280 2.928) \lvec(5.296 2.929) \lvec(5.312
2.929) \lvec(5.328 2.930) \lvec(5.344 2.931) \lvec(5.360 2.931)
\lvec(5.376 2.932) \lvec(5.392 2.933) \lvec(5.408 2.933) \lvec(5.424
2.934) \lvec(5.440 2.934) \lvec(5.456 2.935) \lvec(5.472 2.936)
\lvec(5.488 2.936) \lvec(5.504 2.937) \lvec(5.520 2.937) \lvec(5.536
2.938) \lvec(5.552 2.939) \lvec(5.568 2.939) \lvec(5.584 2.940)
\lvec(5.600 2.940) \lvec(5.616 2.941) \lvec(5.632 2.941) \lvec(5.648
2.942) \lvec(5.664 2.942) \lvec(5.680 2.943) \lvec(5.696 2.943)
\lvec(5.712 2.944) \lvec(5.728 2.944) \lvec(5.744 2.945) \lvec(5.760
2.945) \lvec(5.776 2.946) \lvec(5.792 2.946) \lvec(5.808 2.947)
\lvec(5.824 2.947) \lvec(5.840 2.948) \lvec(5.856 2.948) \lvec(5.872
2.948) \lvec(5.888 2.949) \lvec(5.904 2.949) \lvec(5.920 2.950)
\lvec(5.936 2.950) \lvec(5.952 2.951) \lvec(5.968 2.951) \lvec(5.984
2.951) \lvec(6.000 2.952) \lvec(6.016 2.952) \lvec(6.032 2.953)
\lvec(6.048 2.953) \lvec(6.064 2.954) \lvec(6.080 2.954) \lvec(6.096
2.954) \lvec(6.112 2.955) \lvec(6.128 2.955) \lvec(6.144 2.955)
\lvec(6.160 2.956) \lvec(6.176 2.956) \lvec(6.192 2.956) \lvec(6.208
2.957) \lvec(6.224 2.957) \lvec(6.240 2.958) \lvec(6.256 2.958)
\lvec(6.272 2.958) \lvec(6.288 2.958) \lvec(6.304 2.959) \lvec(6.320
2.959) \lvec(6.336 2.959) \lvec(6.352 2.960) \lvec(6.368 2.960)
\lvec(6.384 2.960) \lvec(6.400 2.961) \lvec(6.416 2.961) \lvec(6.432
2.961) \lvec(6.448 2.962) \lvec(6.464 2.962) \lvec(6.480 2.962)
\lvec(6.496 2.962) \lvec(6.512 2.963) \lvec(6.528 2.963) \lvec(6.544
2.963) \lvec(6.560 2.964) \lvec(6.576 2.964) \lvec(6.592 2.964)
\lvec(6.608 2.964) \lvec(6.624 2.965) \lvec(6.640 2.965) \lvec(6.656
2.965) \lvec(6.672 2.965) \lvec(6.688 2.966) \lvec(6.704 2.966)
\lvec(6.720 2.966) \lvec(6.736 2.966) \lvec(6.752 2.966) \lvec(6.768
2.967) \lvec(6.784 2.967) \lvec(6.800 2.967) \lvec(6.816 2.967)
\lvec(6.832 2.968) \lvec(6.848 2.968) \lvec(6.864 2.968) \lvec(6.880
2.968) \lvec(6.896 2.969) \lvec(6.912 2.969) \lvec(6.928 2.969)
\lvec(6.944 2.969) \lvec(6.960 2.969) \lvec(6.976 2.969) \lvec(6.992
2.970) \lvec(7.008 2.970) \lvec(7.024 2.970) \lvec(7.040 2.970)
\lvec(7.056 2.970) \lvec(7.072 2.971) \lvec(7.088 2.971) \lvec(7.104
2.971) \lvec(7.120 2.971) \lvec(7.136 2.971) \lvec(7.152 2.972)
\lvec(7.168 2.972) \lvec(7.184 2.972) \lvec(7.200 2.972) \lvec(7.216
2.972) \lvec(7.232 2.972) \lvec(7.248 2.973) \lvec(7.264 2.973)
\lvec(7.280 2.973) \lvec(7.296 2.973) \lvec(7.312 2.973) \lvec(7.328
2.973) \lvec(7.344 2.973) \lvec(7.360 2.974) \lvec(7.376 2.974)
\lvec(7.392 2.974) \lvec(7.408 2.974) \lvec(7.424 2.974) \lvec(7.440
2.974) \lvec(7.456 2.975) \lvec(7.472 2.975) \lvec(7.488 2.975)
\lvec(7.504 2.975) \lvec(7.520 2.975) \lvec(7.536 2.975) \lvec(7.552
2.975) \lvec(7.568 2.975) \lvec(7.584 2.976) \lvec(7.600 2.976)
\lvec(7.616 2.976) \lvec(7.632 2.976) \lvec(7.648 2.976) \lvec(7.664
2.976) \lvec(7.680 2.976) \lvec(7.696 2.976) \lvec(7.712 2.976)
\lvec(7.728 2.977) \lvec(7.744 2.977) \lvec(7.760 2.977) \lvec(7.776
2.977) \lvec(7.792 2.977) \lvec(7.808 2.977) \lvec(7.824 2.977)
\lvec(7.840 2.977) \lvec(7.856 2.977) \lvec(7.872 2.978) \lvec(7.888
2.978) \lvec(7.904 2.978) \lvec(7.920 2.978) \lvec(7.936 2.978)
\lvec(7.952 2.978) \lvec(7.968 2.978) \lvec(7.984 2.978) \lvec(8.000
2.978)
%
%
\move(0.000 1.000) \lvec(0.016 1.000) \lvec(0.032 1.000) \lvec(0.048
1.001) \lvec(0.064 1.003) \lvec(0.080 1.006) \lvec(0.096 1.010)
\lvec(0.112 1.014) \lvec(0.128 1.018) \lvec(0.144 1.024) \lvec(0.160
1.029) \lvec(0.176 1.036) \lvec(0.192 1.043) \lvec(0.208 1.050)
\lvec(0.224 1.058) \lvec(0.240 1.066) \lvec(0.256 1.075) \lvec(0.272
1.084) \lvec(0.288 1.093) \lvec(0.304 1.103) \lvec(0.320 1.113)
\lvec(0.336 1.123) \lvec(0.352 1.134) \lvec(0.368 1.144) \lvec(0.384
1.156) \lvec(0.400 1.167) \lvec(0.416 1.179) \lvec(0.432 1.190)
\lvec(0.448 1.202) \lvec(0.464 1.215) \lvec(0.480 1.227) \lvec(0.496
1.239) \lvec(0.512 1.252) \lvec(0.528 1.265) \lvec(0.544 1.277)
\lvec(0.560 1.290) \lvec(0.576 1.303) \lvec(0.592 1.316) \lvec(0.608
1.330) \lvec(0.624 1.343) \lvec(0.640 1.356) \lvec(0.656 1.370)
\lvec(0.672 1.383) \lvec(0.688 1.397) \lvec(0.704 1.410) \lvec(0.720
1.424) \lvec(0.736 1.437) \lvec(0.752 1.451) \lvec(0.768 1.464)
\lvec(0.784 1.478) \lvec(0.800 1.491) \lvec(0.816 1.505) \lvec(0.832
1.518) \lvec(0.848 1.532) \lvec(0.864 1.545) \lvec(0.880 1.558)
\lvec(0.896 1.572) \lvec(0.912 1.585) \lvec(0.928 1.599) \lvec(0.944
1.612) \lvec(0.960 1.625) \lvec(0.976 1.638) \lvec(0.992 1.651)
\lvec(1.008 1.664) \lvec(1.024 1.677) \lvec(1.040 1.690) \lvec(1.056
1.703) \lvec(1.072 1.716) \lvec(1.088 1.728) \lvec(1.104 1.741)
\lvec(1.120 1.754) \lvec(1.136 1.766) \lvec(1.152 1.779) \lvec(1.168
1.791) \lvec(1.184 1.803) \lvec(1.200 1.815) \lvec(1.216 1.827)
\lvec(1.232 1.839) \lvec(1.248 1.851) \lvec(1.264 1.863) \lvec(1.280
1.875) \lvec(1.296 1.887) \lvec(1.312 1.898) \lvec(1.328 1.910)
\lvec(1.344 1.921) \lvec(1.360 1.932) \lvec(1.376 1.944) \lvec(1.392
1.955) \lvec(1.408 1.966) \lvec(1.424 1.977) \lvec(1.440 1.988)
\lvec(1.456 1.998) \lvec(1.472 2.009) \lvec(1.488 2.020) \lvec(1.504
2.030) \lvec(1.520 2.040) \lvec(1.536 2.051) \lvec(1.552 2.061)
\lvec(1.568 2.071) \lvec(1.584 2.081) \lvec(1.600 2.091) \lvec(1.616
2.101) \lvec(1.632 2.111) \lvec(1.648 2.120) \lvec(1.664 2.130)
\lvec(1.680 2.139) \lvec(1.696 2.149) \lvec(1.712 2.158) \lvec(1.728
2.167) \lvec(1.744 2.176) \lvec(1.760 2.185) \lvec(1.776 2.194)
\lvec(1.792 2.203) \lvec(1.808 2.212) \lvec(1.824 2.221) \lvec(1.840
2.229) \lvec(1.856 2.238) \lvec(1.872 2.246) \lvec(1.888 2.254)
\lvec(1.904 2.263) \lvec(1.920 2.271) \lvec(1.936 2.279) \lvec(1.952
2.287) \lvec(1.968 2.295) \lvec(1.984 2.303) \lvec(2.000 2.311)
\lvec(2.016 2.318) \lvec(2.032 2.326) \lvec(2.048 2.333) \lvec(2.064
2.341) \lvec(2.080 2.348) \lvec(2.096 2.355) \lvec(2.112 2.363)
\lvec(2.128 2.370) \lvec(2.144 2.377) \lvec(2.160 2.384) \lvec(2.176
2.391) \lvec(2.192 2.398) \lvec(2.208 2.404) \lvec(2.224 2.411)
\lvec(2.240 2.418) \lvec(2.256 2.424) \lvec(2.272 2.431) \lvec(2.288
2.437) \lvec(2.304 2.444) \lvec(2.320 2.450) \lvec(2.336 2.456)
\lvec(2.352 2.462) \lvec(2.368 2.468) \lvec(2.384 2.474) \lvec(2.400
2.480) \lvec(2.416 2.486) \lvec(2.432 2.492) \lvec(2.448 2.497)
\lvec(2.464 2.503) \lvec(2.480 2.509) \lvec(2.496 2.514) \lvec(2.512
2.520) \lvec(2.528 2.525) \lvec(2.544 2.531) \lvec(2.560 2.536)
\lvec(2.576 2.541) \lvec(2.592 2.546) \lvec(2.608 2.551) \lvec(2.624
2.556) \lvec(2.640 2.562) \lvec(2.656 2.566) \lvec(2.672 2.571)
\lvec(2.688 2.576) \lvec(2.704 2.581) \lvec(2.720 2.586) \lvec(2.736
2.590) \lvec(2.752 2.595) \lvec(2.768 2.600) \lvec(2.784 2.604)
\lvec(2.800 2.609) \lvec(2.816 2.613) \lvec(2.832 2.617) \lvec(2.848
2.622) \lvec(2.864 2.626) \lvec(2.880 2.630) \lvec(2.896 2.634)
\lvec(2.912 2.638) \lvec(2.928 2.643) \lvec(2.944 2.647) \lvec(2.960
2.651) \lvec(2.976 2.655) \lvec(2.992 2.658) \lvec(3.008 2.662)
\lvec(3.024 2.666) \lvec(3.040 2.670) \lvec(3.056 2.674) \lvec(3.072
2.677) \lvec(3.088 2.681) \lvec(3.104 2.685) \lvec(3.120 2.688)
\lvec(3.136 2.692) \lvec(3.152 2.695) \lvec(3.168 2.698) \lvec(3.184
2.702) \lvec(3.200 2.705) \lvec(3.216 2.708) \lvec(3.232 2.712)
\lvec(3.248 2.715) \lvec(3.264 2.718) \lvec(3.280 2.721) \lvec(3.296
2.724) \lvec(3.312 2.728) \lvec(3.328 2.731) \lvec(3.344 2.734)
\lvec(3.360 2.737) \lvec(3.376 2.740) \lvec(3.392 2.742) \lvec(3.408
2.745) \lvec(3.424 2.748) \lvec(3.440 2.751) \lvec(3.456 2.754)
\lvec(3.472 2.757) \lvec(3.488 2.759) \lvec(3.504 2.762) \lvec(3.520
2.765) \lvec(3.536 2.767) \lvec(3.552 2.770) \lvec(3.568 2.772)
\lvec(3.584 2.775) \lvec(3.600 2.777) \lvec(3.616 2.780) \lvec(3.632
2.782) \lvec(3.648 2.785) \lvec(3.664 2.787) \lvec(3.680 2.789)
\lvec(3.696 2.792) \lvec(3.712 2.794) \lvec(3.728 2.796) \lvec(3.744
2.799) \lvec(3.760 2.801) \lvec(3.776 2.803) \lvec(3.792 2.805)
\lvec(3.808 2.807) \lvec(3.824 2.810) \lvec(3.840 2.811) \lvec(3.856
2.814) \lvec(3.872 2.816) \lvec(3.888 2.818) \lvec(3.904 2.820)
\lvec(3.920 2.822) \lvec(3.936 2.824) \lvec(3.952 2.825) \lvec(3.968
2.827) \lvec(3.984 2.829) \lvec(4.000 2.831) \lvec(4.016 2.833)
\lvec(4.032 2.835) \lvec(4.048 2.837) \lvec(4.064 2.838) \lvec(4.080
2.840) \lvec(4.096 2.842) \lvec(4.112 2.844) \lvec(4.128 2.845)
\lvec(4.144 2.847) \lvec(4.160 2.849) \lvec(4.176 2.850) \lvec(4.192
2.852) \lvec(4.208 2.854) \lvec(4.224 2.855) \lvec(4.240 2.857)
\lvec(4.256 2.858) \lvec(4.272 2.860) \lvec(4.288 2.861) \lvec(4.304
2.863) \lvec(4.320 2.864) \lvec(4.336 2.866) \lvec(4.352 2.867)
\lvec(4.368 2.868) \lvec(4.384 2.870) \lvec(4.400 2.871) \lvec(4.416
2.873) \lvec(4.432 2.874) \lvec(4.448 2.875) \lvec(4.464 2.877)
\lvec(4.480 2.878) \lvec(4.496 2.879) \lvec(4.512 2.880) \lvec(4.528
2.882) \lvec(4.544 2.883) \lvec(4.560 2.884) \lvec(4.576 2.885)
\lvec(4.592 2.887) \lvec(4.608 2.888) \lvec(4.624 2.889) \lvec(4.640
2.890) \lvec(4.656 2.891) \lvec(4.672 2.892) \lvec(4.688 2.893)
\lvec(4.704 2.895) \lvec(4.720 2.896) \lvec(4.736 2.897) \lvec(4.752
2.898) \lvec(4.768 2.899) \lvec(4.784 2.900) \lvec(4.800 2.901)
\lvec(4.816 2.902) \lvec(4.832 2.903) \lvec(4.848 2.904) \lvec(4.864
2.905) \lvec(4.880 2.906) \lvec(4.896 2.907) \lvec(4.912 2.908)
\lvec(4.928 2.909) \lvec(4.944 2.910) \lvec(4.960 2.911) \lvec(4.976
2.911) \lvec(4.992 2.912) \lvec(5.008 2.913) \lvec(5.024 2.914)
\lvec(5.040 2.915) \lvec(5.056 2.916) \lvec(5.072 2.917) \lvec(5.088
2.917) \lvec(5.104 2.918) \lvec(5.120 2.919) \lvec(5.136 2.920)
\lvec(5.152 2.921) \lvec(5.168 2.921) \lvec(5.184 2.922) \lvec(5.200
2.923) \lvec(5.216 2.924) \lvec(5.232 2.924) \lvec(5.248 2.925)
\lvec(5.264 2.926) \lvec(5.280 2.927) \lvec(5.296 2.927) \lvec(5.312
2.928) \lvec(5.328 2.929) \lvec(5.344 2.929) \lvec(5.360 2.930)
\lvec(5.376 2.931) \lvec(5.392 2.931) \lvec(5.408 2.932) \lvec(5.424
2.933) \lvec(5.440 2.933) \lvec(5.456 2.934) \lvec(5.472 2.935)
\lvec(5.488 2.935) \lvec(5.504 2.936) \lvec(5.520 2.936) \lvec(5.536
2.937) \lvec(5.552 2.938) \lvec(5.568 2.938) \lvec(5.584 2.939)
\lvec(5.600 2.939) \lvec(5.616 2.940) \lvec(5.632 2.940) \lvec(5.648
2.941) \lvec(5.664 2.941) \lvec(5.680 2.942) \lvec(5.696 2.943)
\lvec(5.712 2.943) \lvec(5.728 2.944) \lvec(5.744 2.944) \lvec(5.760
2.945) \lvec(5.776 2.945) \lvec(5.792 2.946) \lvec(5.808 2.946)
\lvec(5.824 2.947) \lvec(5.840 2.947) \lvec(5.856 2.947) \lvec(5.872
2.948) \lvec(5.888 2.948) \lvec(5.904 2.949) \lvec(5.920 2.949)
\lvec(5.936 2.950) \lvec(5.952 2.950) \lvec(5.968 2.951) \lvec(5.984
2.951) \lvec(6.000 2.951) \lvec(6.016 2.952) \lvec(6.032 2.952)
\lvec(6.048 2.953) \lvec(6.064 2.953) \lvec(6.080 2.953) \lvec(6.096
2.954) \lvec(6.112 2.954) \lvec(6.128 2.955) \lvec(6.144 2.955)
\lvec(6.160 2.955) \lvec(6.176 2.956) \lvec(6.192 2.956) \lvec(6.208
2.956) \lvec(6.224 2.957) \lvec(6.240 2.957) \lvec(6.256 2.958)
\lvec(6.272 2.958) \lvec(6.288 2.958) \lvec(6.304 2.958) \lvec(6.320
2.959) \lvec(6.336 2.959) \lvec(6.352 2.959) \lvec(6.368 2.960)
\lvec(6.384 2.960) \lvec(6.400 2.960) \lvec(6.416 2.961) \lvec(6.432
2.961) \lvec(6.448 2.961) \lvec(6.464 2.962) \lvec(6.480 2.962)
\lvec(6.496 2.962) \lvec(6.512 2.962) \lvec(6.528 2.963) \lvec(6.544
2.963) \lvec(6.560 2.963) \lvec(6.576 2.964) \lvec(6.592 2.964)
\lvec(6.608 2.964) \lvec(6.624 2.964) \lvec(6.640 2.965) \lvec(6.656
2.965) \lvec(6.672 2.965) \lvec(6.688 2.965) \lvec(6.704 2.966)
\lvec(6.720 2.966) \lvec(6.736 2.966) \lvec(6.752 2.966) \lvec(6.768
2.967) \lvec(6.784 2.967) \lvec(6.800 2.967) \lvec(6.816 2.967)
\lvec(6.832 2.968) \lvec(6.848 2.968) \lvec(6.864 2.968) \lvec(6.880
2.968) \lvec(6.896 2.968) \lvec(6.912 2.969) \lvec(6.928 2.969)
\lvec(6.944 2.969) \lvec(6.960 2.969) \lvec(6.976 2.969) \lvec(6.992
2.970) \lvec(7.008 2.970) \lvec(7.024 2.970) \lvec(7.040 2.970)
\lvec(7.056 2.970) \lvec(7.072 2.971) \lvec(7.088 2.971) \lvec(7.104
2.971) \lvec(7.120 2.971) \lvec(7.136 2.971) \lvec(7.152 2.972)
\lvec(7.168 2.972) \lvec(7.184 2.972) \lvec(7.200 2.972) \lvec(7.216
2.972) \lvec(7.232 2.972) \lvec(7.248 2.972) \lvec(7.264 2.973)
\lvec(7.280 2.973) \lvec(7.296 2.973) \lvec(7.312 2.973) \lvec(7.328
2.973) \lvec(7.344 2.973) \lvec(7.360 2.974) \lvec(7.376 2.974)
\lvec(7.392 2.974) \lvec(7.408 2.974) \lvec(7.424 2.974) \lvec(7.440
2.974) \lvec(7.456 2.975) \lvec(7.472 2.975) \lvec(7.488 2.975)
\lvec(7.504 2.975) \lvec(7.520 2.975) \lvec(7.536 2.975) \lvec(7.552
2.975) \lvec(7.568 2.975) \lvec(7.584 2.976) \lvec(7.600 2.976)
\lvec(7.616 2.976) \lvec(7.632 2.976) \lvec(7.648 2.976) \lvec(7.664
2.976) \lvec(7.680 2.976) \lvec(7.696 2.976) \lvec(7.712 2.976)
\lvec(7.728 2.977) \lvec(7.744 2.977) \lvec(7.760 2.977) \lvec(7.776
2.977) \lvec(7.792 2.977) \lvec(7.808 2.977) \lvec(7.824 2.977)
\lvec(7.840 2.977) \lvec(7.856 2.977) \lvec(7.872 2.978) \lvec(7.888
2.978) \lvec(7.904 2.978) \lvec(7.920 2.978) \lvec(7.936 2.978)
\lvec(7.952 2.978) \lvec(7.968 2.978) \lvec(7.984 2.978) \lvec(8.000
2.978)
\move(0.000 3.000) \lvec(0.016 2.960) \lvec(0.032 2.921) \lvec(0.048
2.884) \lvec(0.064 2.848) \lvec(0.080 2.814) \lvec(0.096 2.781)
\lvec(0.112 2.749) \lvec(0.128 2.718) \lvec(0.144 2.689) \lvec(0.160
2.660) \lvec(0.176 2.633) \lvec(0.192 2.607) \lvec(0.208 2.582)
\lvec(0.224 2.558) \lvec(0.240 2.535) \lvec(0.256 2.513) \lvec(0.272
2.492) \lvec(0.288 2.472) \lvec(0.304 2.452) \lvec(0.320 2.434)
\lvec(0.336 2.416) \lvec(0.352 2.399) \lvec(0.368 2.383) \lvec(0.384
2.368) \lvec(0.400 2.354) \lvec(0.416 2.340) \lvec(0.432 2.327)
\lvec(0.448 2.314) \lvec(0.464 2.302) \lvec(0.480 2.291) \lvec(0.496
2.281) \lvec(0.512 2.271) \lvec(0.528 2.261) \lvec(0.544 2.252)
\lvec(0.560 2.244) \lvec(0.576 2.236) \lvec(0.592 2.229) \lvec(0.608
2.222) \lvec(0.624 2.216) \lvec(0.640 2.210) \lvec(0.656 2.204)
\lvec(0.672 2.199) \lvec(0.688 2.195) \lvec(0.704 2.191) \lvec(0.720
2.187) \lvec(0.736 2.183) \lvec(0.752 2.180) \lvec(0.768 2.178)
\lvec(0.784 2.175) \lvec(0.800 2.173) \lvec(0.816 2.171) \lvec(0.832
2.170) \lvec(0.848 2.169) \lvec(0.864 2.168) \lvec(0.880 2.167)
\lvec(0.896 2.167) \lvec(0.912 2.167) \lvec(0.928 2.167) \lvec(0.944
2.167) \lvec(0.960 2.168) \lvec(0.976 2.169) \lvec(0.992 2.170)
\lvec(1.008 2.171) \lvec(1.024 2.172) \lvec(1.040 2.174) \lvec(1.056
2.176) \lvec(1.072 2.178) \lvec(1.088 2.180) \lvec(1.104 2.182)
\lvec(1.120 2.185) \lvec(1.136 2.187) \lvec(1.152 2.190) \lvec(1.168
2.193) \lvec(1.184 2.196) \lvec(1.200 2.199) \lvec(1.216 2.202)
\lvec(1.232 2.206) \lvec(1.248 2.209) \lvec(1.264 2.213) \lvec(1.280
2.216) \lvec(1.296 2.220) \lvec(1.312 2.224) \lvec(1.328 2.228)
\lvec(1.344 2.232) \lvec(1.360 2.236) \lvec(1.376 2.240) \lvec(1.392
2.244) \lvec(1.408 2.249) \lvec(1.424 2.253) \lvec(1.440 2.258)
\lvec(1.456 2.262) \lvec(1.472 2.266) \lvec(1.488 2.271) \lvec(1.504
2.276) \lvec(1.520 2.280) \lvec(1.536 2.285) \lvec(1.552 2.290)
\lvec(1.568 2.294) \lvec(1.584 2.299) \lvec(1.600 2.304) \lvec(1.616
2.309) \lvec(1.632 2.314) \lvec(1.648 2.319) \lvec(1.664 2.323)
\lvec(1.680 2.329) \lvec(1.696 2.333) \lvec(1.712 2.338) \lvec(1.728
2.343) \lvec(1.744 2.348) \lvec(1.760 2.353) \lvec(1.776 2.358)
\lvec(1.792 2.363) \lvec(1.808 2.368) \lvec(1.824 2.373) \lvec(1.840
2.378) \lvec(1.856 2.383) \lvec(1.872 2.388) \lvec(1.888 2.393)
\lvec(1.904 2.398) \lvec(1.920 2.403) \lvec(1.936 2.408) \lvec(1.952
2.413) \lvec(1.968 2.418) \lvec(1.984 2.423) \lvec(2.000 2.428)
\lvec(2.016 2.433) \lvec(2.032 2.437) \lvec(2.048 2.442) \lvec(2.064
2.447) \lvec(2.080 2.452) \lvec(2.096 2.457) \lvec(2.112 2.462)
\lvec(2.128 2.466) \lvec(2.144 2.471) \lvec(2.160 2.476) \lvec(2.176
2.481) \lvec(2.192 2.485) \lvec(2.208 2.490) \lvec(2.224 2.495)
\lvec(2.240 2.499) \lvec(2.256 2.504) \lvec(2.272 2.508) \lvec(2.288
2.513) \lvec(2.304 2.517) \lvec(2.320 2.522) \lvec(2.336 2.526)
\lvec(2.352 2.531) \lvec(2.368 2.535) \lvec(2.384 2.540) \lvec(2.400
2.544) \lvec(2.416 2.548) \lvec(2.432 2.553) \lvec(2.448 2.557)
\lvec(2.464 2.561) \lvec(2.480 2.565) \lvec(2.496 2.570) \lvec(2.512
2.574) \lvec(2.528 2.578) \lvec(2.544 2.582) \lvec(2.560 2.586)
\lvec(2.576 2.590) \lvec(2.592 2.594) \lvec(2.608 2.598) \lvec(2.624
2.602) \lvec(2.640 2.606) \lvec(2.656 2.610) \lvec(2.672 2.614)
\lvec(2.688 2.617) \lvec(2.704 2.621) \lvec(2.720 2.625) \lvec(2.736
2.629) \lvec(2.752 2.632) \lvec(2.768 2.636) \lvec(2.784 2.640)
\lvec(2.800 2.643) \lvec(2.816 2.647) \lvec(2.832 2.650) \lvec(2.848
2.654) \lvec(2.864 2.657) \lvec(2.880 2.661) \lvec(2.896 2.664)
\lvec(2.912 2.668) \lvec(2.928 2.671) \lvec(2.944 2.674) \lvec(2.960
2.678) \lvec(2.976 2.681) \lvec(2.992 2.684) \lvec(3.008 2.688)
\lvec(3.024 2.691) \lvec(3.040 2.694) \lvec(3.056 2.697) \lvec(3.072
2.700) \lvec(3.088 2.703) \lvec(3.104 2.706) \lvec(3.120 2.709)
\lvec(3.136 2.712) \lvec(3.152 2.715) \lvec(3.168 2.718) \lvec(3.184
2.721) \lvec(3.200 2.724) \lvec(3.216 2.727) \lvec(3.232 2.730)
\lvec(3.248 2.732) \lvec(3.264 2.735) \lvec(3.280 2.738) \lvec(3.296
2.741) \lvec(3.312 2.743) \lvec(3.328 2.746) \lvec(3.344 2.749)
\lvec(3.360 2.751) \lvec(3.376 2.754) \lvec(3.392 2.757) \lvec(3.408
2.759) \lvec(3.424 2.762) \lvec(3.440 2.764) \lvec(3.456 2.766)
\lvec(3.472 2.769) \lvec(3.488 2.771) \lvec(3.504 2.774) \lvec(3.520
2.776) \lvec(3.536 2.778) \lvec(3.552 2.781) \lvec(3.568 2.783)
\lvec(3.584 2.785) \lvec(3.600 2.788) \lvec(3.616 2.790) \lvec(3.632
2.792) \lvec(3.648 2.794) \lvec(3.664 2.796) \lvec(3.680 2.798)
\lvec(3.696 2.801) \lvec(3.712 2.803) \lvec(3.728 2.805) \lvec(3.744
2.807) \lvec(3.760 2.809) \lvec(3.776 2.811) \lvec(3.792 2.813)
\lvec(3.808 2.815) \lvec(3.824 2.817) \lvec(3.840 2.819) \lvec(3.856
2.821) \lvec(3.872 2.822) \lvec(3.888 2.824) \lvec(3.904 2.826)
\lvec(3.920 2.828) \lvec(3.936 2.830) \lvec(3.952 2.831) \lvec(3.968
2.833) \lvec(3.984 2.835) \lvec(4.000 2.837) \lvec(4.016 2.838)
\lvec(4.032 2.840) \lvec(4.048 2.842) \lvec(4.064 2.843) \lvec(4.080
2.845) \lvec(4.096 2.847) \lvec(4.112 2.848) \lvec(4.128 2.850)
\lvec(4.144 2.851) \lvec(4.160 2.853) \lvec(4.176 2.854) \lvec(4.192
2.856) \lvec(4.208 2.857) \lvec(4.224 2.859) \lvec(4.240 2.860)
\lvec(4.256 2.862) \lvec(4.272 2.863) \lvec(4.288 2.865) \lvec(4.304
2.866) \lvec(4.320 2.868) \lvec(4.336 2.869) \lvec(4.352 2.870)
\lvec(4.368 2.872) \lvec(4.384 2.873) \lvec(4.400 2.874) \lvec(4.416
2.875) \lvec(4.432 2.877) \lvec(4.448 2.878) \lvec(4.464 2.879)
\lvec(4.480 2.880) \lvec(4.496 2.882) \lvec(4.512 2.883) \lvec(4.528
2.884) \lvec(4.544 2.885) \lvec(4.560 2.886) \lvec(4.576 2.888)
\lvec(4.592 2.889) \lvec(4.608 2.890) \lvec(4.624 2.891) \lvec(4.640
2.892) \lvec(4.656 2.893) \lvec(4.672 2.894) \lvec(4.688 2.895)
\lvec(4.704 2.896) \lvec(4.720 2.897) \lvec(4.736 2.898) \lvec(4.752
2.900) \lvec(4.768 2.901) \lvec(4.784 2.901) \lvec(4.800 2.903)
\lvec(4.816 2.904) \lvec(4.832 2.904) \lvec(4.848 2.905) \lvec(4.864
2.906) \lvec(4.880 2.907) \lvec(4.896 2.908) \lvec(4.912 2.909)
\lvec(4.928 2.910) \lvec(4.944 2.911) \lvec(4.960 2.912) \lvec(4.976
2.913) \lvec(4.992 2.914) \lvec(5.008 2.914) \lvec(5.024 2.915)
\lvec(5.040 2.916) \lvec(5.056 2.917) \lvec(5.072 2.918) \lvec(5.088
2.918) \lvec(5.104 2.919) \lvec(5.120 2.920) \lvec(5.136 2.921)
\lvec(5.152 2.922) \lvec(5.168 2.922) \lvec(5.184 2.923) \lvec(5.200
2.924) \lvec(5.216 2.924) \lvec(5.232 2.925) \lvec(5.248 2.926)
\lvec(5.264 2.927) \lvec(5.280 2.927) \lvec(5.296 2.928) \lvec(5.312
2.929) \lvec(5.328 2.929) \lvec(5.344 2.930) \lvec(5.360 2.931)
\lvec(5.376 2.931) \lvec(5.392 2.932) \lvec(5.408 2.933) \lvec(5.424
2.933) \lvec(5.440 2.934) \lvec(5.456 2.934) \lvec(5.472 2.935)
\lvec(5.488 2.936) \lvec(5.504 2.936) \lvec(5.520 2.937) \lvec(5.536
2.938) \lvec(5.552 2.938) \lvec(5.568 2.939) \lvec(5.584 2.939)
\lvec(5.600 2.940) \lvec(5.616 2.940) \lvec(5.632 2.941) \lvec(5.648
2.941) \lvec(5.664 2.942) \lvec(5.680 2.942) \lvec(5.696 2.943)
\lvec(5.712 2.944) \lvec(5.728 2.944) \lvec(5.744 2.944) \lvec(5.760
2.945) \lvec(5.776 2.945) \lvec(5.792 2.946) \lvec(5.808 2.946)
\lvec(5.824 2.947) \lvec(5.840 2.947) \lvec(5.856 2.948) \lvec(5.872
2.948) \lvec(5.888 2.949) \lvec(5.904 2.949) \lvec(5.920 2.950)
\lvec(5.936 2.950) \lvec(5.952 2.950) \lvec(5.968 2.951) \lvec(5.984
2.951) \lvec(6.000 2.952) \lvec(6.016 2.952) \lvec(6.032 2.952)
\lvec(6.048 2.953) \lvec(6.064 2.953) \lvec(6.080 2.954) \lvec(6.096
2.954) \lvec(6.112 2.954) \lvec(6.128 2.955) \lvec(6.144 2.955)
\lvec(6.160 2.955) \lvec(6.176 2.956) \lvec(6.192 2.956) \lvec(6.208
2.957) \lvec(6.224 2.957) \lvec(6.240 2.957) \lvec(6.256 2.958)
\lvec(6.272 2.958) \lvec(6.288 2.958) \lvec(6.304 2.959) \lvec(6.320
2.959) \lvec(6.336 2.959) \lvec(6.352 2.960) \lvec(6.368 2.960)
\lvec(6.384 2.960) \lvec(6.400 2.961) \lvec(6.416 2.961) \lvec(6.432
2.961) \lvec(6.448 2.961) \lvec(6.464 2.962) \lvec(6.480 2.962)
\lvec(6.496 2.962) \lvec(6.512 2.963) \lvec(6.528 2.963) \lvec(6.544
2.963) \lvec(6.560 2.963) \lvec(6.576 2.964) \lvec(6.592 2.964)
\lvec(6.608 2.964) \lvec(6.624 2.965) \lvec(6.640 2.965) \lvec(6.656
2.965) \lvec(6.672 2.965) \lvec(6.688 2.965) \lvec(6.704 2.966)
\lvec(6.720 2.966) \lvec(6.736 2.966) \lvec(6.752 2.966) \lvec(6.768
2.967) \lvec(6.784 2.967) \lvec(6.800 2.967) \lvec(6.816 2.967)
\lvec(6.832 2.968) \lvec(6.848 2.968) \lvec(6.864 2.968) \lvec(6.880
2.968) \lvec(6.896 2.968) \lvec(6.912 2.969) \lvec(6.928 2.969)
\lvec(6.944 2.969) \lvec(6.960 2.969) \lvec(6.976 2.969) \lvec(6.992
2.970) \lvec(7.008 2.970) \lvec(7.024 2.970) \lvec(7.040 2.970)
\lvec(7.056 2.970) \lvec(7.072 2.971) \lvec(7.088 2.971) \lvec(7.104
2.971) \lvec(7.120 2.971) \lvec(7.136 2.971) \lvec(7.152 2.972)
\lvec(7.168 2.972) \lvec(7.184 2.972) \lvec(7.200 2.972) \lvec(7.216
2.972) \lvec(7.232 2.972) \lvec(7.248 2.972) \lvec(7.264 2.973)
\lvec(7.280 2.973) \lvec(7.296 2.973) \lvec(7.312 2.973) \lvec(7.328
2.973) \lvec(7.344 2.973) \lvec(7.360 2.974) \lvec(7.376 2.974)
\lvec(7.392 2.974) \lvec(7.408 2.974) \lvec(7.424 2.974) \lvec(7.440
2.974) \lvec(7.456 2.975) \lvec(7.472 2.975) \lvec(7.488 2.975)
\lvec(7.504 2.975) \lvec(7.520 2.975) \lvec(7.536 2.975) \lvec(7.552
2.975) \lvec(7.568 2.975) \lvec(7.584 2.976) \lvec(7.600 2.976)
\lvec(7.616 2.976) \lvec(7.632 2.976) \lvec(7.648 2.976) \lvec(7.664
2.976) \lvec(7.680 2.976) \lvec(7.696 2.976) \lvec(7.712 2.976)
\lvec(7.728 2.977) \lvec(7.744 2.977) \lvec(7.760 2.977) \lvec(7.776
2.977) \lvec(7.792 2.977) \lvec(7.808 2.977) \lvec(7.824 2.977)
\lvec(7.840 2.977) \lvec(7.856 2.977) \lvec(7.872 2.978) \lvec(7.888
2.978) \lvec(7.904 2.978) \lvec(7.920 2.978) \lvec(7.936 2.978)
\lvec(7.952 2.978) \lvec(7.968 2.978) \lvec(7.984 2.978) \lvec(8.000
2.978)
\end{texdraw}
\end{center}
%
%
\caption{Boundary layers in the mean degree of orientation
$<s(x,z)>_x$, when $\xi=0.25 \eta$, $\spr=0.8$, $\omega=0.6$,
$\mr=0.1\eta$, and $\Delta=1.5$, when Dirichlet-like boundary
conditions are applied on the degree of orientation. The boundary
degree of orientation $\tilde s$ is respectively equal to 1 (top),
$\spr$ (middle), and $0.6$ (bottom).} \label{s_sol_dir}
\end{figure}

Figure \ref{s_sol_dir} shows how the degree of orientation varies
within the boundary layer, as $\tilde s$ is fixed above, equal to,
or below $\spr$. A double boundary-layer structure emerges. All
plots exhibit a decrease of $s$ in a region of characteristic size
$\eta$: this effect comes from the $\BO{\eps^2}$-contribution. A
similar surface melting was already presented and discussed in
Figure \ref{s_sol_neu_gr1}. Close to the boundary, the $\BO{1}$-term
proportional to $(\tilde s-\spr)\, \er^{-\sqrt{3}\,\omega z/\xi}$
settles the desired boundary value of $s$ in a thin boundary layer
of characteristic size $\xi$.

\section{Effective weak anchoring}\label{sec:effwa}

Once the boundary layer effects fade away, the main macroscopic
effect of a rough surface on the director orientation is to allow
for an effective surface tilt angle $\theta_{\rm b}(0)$, that
apparently violates the homeotropic prescription $\theta(0)=0$ (see
(\ref{thetapp}) and (\ref{ttdir0})). It appears then natural to
check whether the same macroscopic effect may be modeled through a
weak anchoring potential, acting on a smooth surface. In this
section we pursue this similarity, and we derive a relation
connecting the microscopic roughness parameters with a macroscopic
anchoring strength.

To solve the weak-anchoring problem, we consider a nematic liquid
crystal which still spreads in the half-space $\mathcal{B}=\{z\geq
0\}$. To better compare our results with classical weak-anchoring
models, we settle within Frank's director theory, and thus look for
the equilibrium distribution that minimizes the free-energy
functional
\begin{equation}
\mathcal{F}[\bv{n}]:=K\int_{\mathcal{B}} \big|\bv{\nabla
n}\big|^2\,dv+W\int_{\partial\mathcal{B}} f_{\rm w}[\bv{n}]\,da\;.
\label{effeQ}
\end{equation}
The bulk free-energy density in the functional (\ref{effeQ}) can be
derived from its order-tensor theory counterpart by setting $s\equiv
1$ in (\ref{fb}). The anchoring potential $f_{\rm w}$ is required to
attain its minimum at the homeotropic anchoring
$\bv{n}\big|_{\partial\mathcal{B}}=\bv{e}_z$, while $W$ is the
\emph{anchoring strength}.

We look again for equilibrium distributions of the type
$\bv{n}(z)=\sin\theta(z)\,\bv{e}_x+\cos\theta(z)\,\bv{e}_z$. Thus,
the free energy functional (\ref{effeQ}) per unit trasverse area can
be written as
\begin{equation}
f[\theta]:=K\int \theta^{\prime 2}(z)\,dz+W\,f_{\rm
w}\big(\theta(0)\big)\;, \label{effett}
\end{equation}
where we assume $f'_{\rm w}(0)=0$ and $f''_{\rm w}(0)>0$, in order
to guarantee the homeotropic preference. The minimizers of
(\ref{effett}) satisfy the trivial Euler-Lagrange equation
$\theta''=0$, and the boundary condition
\begin{equation}
K\theta'(0)-Wf'_{\rm w}\big(\theta(0)\big)=0\;.\label{bc3}
\end{equation}
When the anchoring strength $W$ is large enough, the boundary
condition (\ref{bc3}) requires $\theta(0)$ to be small. When this is
the case, a Taylor expansion in (\ref{bc3}) supplies
\begin{equation}
\theta(0)\approx \frac{K\,\mr}{Wf''_{\rm w}(0)} =
\lambda\,\mr\;,\label{bcris}
\end{equation}
In (\ref{bcris}) we have restored the notation $\mr=\theta'(0)$ to
better compare this estimate with our preceding results, and
introduced the \emph{surface extrapolation length}
\begin{equation}
\lambda:=\frac{K}{Wf''_{\rm w}(0)}\;,\label{exl}
\end{equation}
a quantity that compares the relative strengths of the elastic and
anchoring potentials.

The comparison between (\ref{bcris}) and our results
(\ref{thetapp})-(\ref{ttdir0}) relates the surface extrapolation
length to the microscopic roughness parameters and/or the surface
value of the degree of orientation. This analogy will be examined in
the following section.

\section{Discussion}\label{sec:disc}

We have examined both the boundary layer structure and the bulk
effects of a rough surface bounding a nematic liquid crystal. Our
main results may be summarized as follows.

\begin{itemize}

\item The roughness of the surface has been modeled by an oscillating
anchoring condition, characterized by an oscillation amplitude
$\Delta$ and a wave length $\eta$. Figures \ref{s_sol_neu_gr1} and
\ref{s_sol_dir} show that the rough boundary induces a partial
melting in a neighborhood (of size $\eta$) of the external boundary.
When Neumann-like boundary conditions are imposed on the degree of
orientation, equation (\ref{ssurf}) quantifies the mean degree of
order at the boundary. On the contrary, were $s$ be forced to a
prescribed value $\tilde s$ on the surface, equations (\ref{sdir1})
and (\ref{sdir2}) show that the boundary condition induces a thin
boundary layer, determined by the nematic coherence length $\xi$.

\item Once the degree of orientation decreases, the spatial
variations of the tilt angle become cheaper, and thus the $\theta$
is keen to steepen close to the external boundary. Figure
\ref{theta_sol_neu_gr1} illustrates this effect. As a consequence,
the effective boundary tilt angle $\theta_{\rm b}(0)$, extrapolated
from the asymptotic outer solution $\theta_{\rm b}(z)$, becomes
different from the null homeotropic prescription (see equations
(\ref{thetapp}) and (\ref{ttdir0})). In the preceding section
\ref{sec:effwa} we have shown that a similar effective anchoring
breaking takes place when a weak anchoring potential is assumed on a
smooth surface (see equation (\ref{exl}) for the characteristic
surface extrapolation length). To further pursue this similarity we
need to consider separately the different anchorings that may be
applied on the degree of orientation.

\begin{itemize}

\item When $s$ is free to choose its boundary value, equation
(\ref{thetapp}) shows that the surface extrapolation length is given
by
\begin{equation}\label{ris1}
\frac{\lambda}{\xi}=\frac{2\Delta^2}{\omega^2}\;\frac{\xi}{\eta}+\mathcal{O}\left(
\frac{\xi^2}{\eta^2}\right)\;.
\end{equation}
Thus, the anchoring strength increases when either the roughness
amplitude $\Delta$ decreases (towards a smooth surface) or the
roughness wave-length increases. An estimate of the order of
magnitude of the effective roughness wave-length can be obtained by
assuming typical values for the quantities involved in (\ref{ris1}).
Indeed, if we assume $\lambda\approx \xi$, $\Delta\approx1$, and
$\omega\approx \frac{1}{2}$ we arrive at $\eta\approx 10\xi$, that
models a roughness wave length in the hundredths of molecular
lengths.

\item When the boundary conditions fix the value of the degree of
orientation at the surface, equation (\ref{ttdir0}) yields
\begin{equation}\label{ris2}
\frac{\lambda}{\xi}=\frac{1}{\sqrt{3}\,\omega}\left(\log
\frac{\spr}{\tilde s} + \frac{\spr-\tilde s}{\tilde s} \right)
+\frac{2\Delta^2}{\omega^2}\;\frac{\xi}{\eta}+\mathcal{O}\left(
\frac{\xi^2}{\eta^2}\right)\;.
\end{equation}
Equation (\ref{ris2}) shows that the surface extrapolation length
includes two quite different contributions. The former depends on
the difference between the boundary and the preferred values of the
degree of orientation ($\tilde s$ and $\spr$, respectively), while
the latter depends on the surface roughness and indeed coincides
with (\ref{ris1}). However, equation (\ref{ris2}) may lose sense
when $\tilde s> \spr$. Indeed, in this case $\lambda$ may become
negative, so providing an \emph{inverse\/} weak anchoring effect.
The physical origin of this odd result may be easily understood if
we again resort to the $s^2|\nabla\theta|^2$-term in the free-energy
density. By virtue of that term, the tilt angle prefers to limit its
spatial variations in regions of higher $s$. If we force in the
surface a higher degree of orientation than the bulk value, the tilt
angle will flatten close to the surface, thus exhibiting the
opposite behaviour with respect to that shown in Figure
\ref{theta_sol_neu_gr1}. Equation (\ref{ris2}) shows that this
inverse effect may occur whenever
\begin{equation}\label{ris3}
\frac{\tilde s - \spr}{\spr} \gtrsim
\frac{\sqrt{3}\Delta^2}{\omega}\;\frac{\xi}{\eta}+\mathcal{O}\left(
\frac{\xi^2}{\eta^2}\right)\;.
\end{equation}
If we again replace the estimates above for $\Delta,\omega,\eta$, we
arrive at the result that a fixed degree of orientation is able to
completely hide the roughness-induced effective weak anchoring
whenever $\tilde s$ exceeds $\spr$ by the 10\% of the preferred
value $\spr$ itself.

\end{itemize}

\end{itemize}

\section*{Acknowledgements}

P.B.\ thanks Georges E.~Durand for useful discussions on the present
topics.

\end{document}